\documentclass[aap,preprint]{imsart} % Also adjusted imsart.sty to remove header

\arxiv{arXiv:1506.06998}

\usepackage{amsthm,amsmath,amssymb,enumerate}
\usepackage[authoryear]{natbib}
\usepackage[ruled, lined, linesnumbered]{algorithm2e} % Must load after natbib! Else strange errors.
\usepackage{mathdots, mathtools}
\usepackage{multicol, multirow}
\startlocaldefs

\newtheorem{lemma}{Lemma}

\newtheorem{proposition}{Proposition}
\newtheorem{theorem}{Theorem}

\newcommand{\sref}[1]{Section~\ref{#1}}
\newcommand{\tref}[1]{Table~\ref{#1}}
\newcommand{\fref}[1]{Figure~\ref{#1}}
\newcommand{\cref}[1]{Chapter~\ref{#1}}

\newcommand{\lemmaref}[1]{Lemma~\ref{#1}}
\newcommand{\propref}[1]{Proposition~\ref{#1}}
\newcommand{\thmref}[1]{Theorem~\ref{#1}}

\def\ds{\ensuremath{\displaystyle}}

\def\bbE{\mathbb{E}}
\def\bbF{\mathbb{F}}

\def\bbN{\mathbb{N}}

\def\bbP{\mathbb{P}}
\def\bbQ{\mathbb{Q}}
\def\bbR{\mathbb{R}}

\def\bbW{\mathbb{W}}

\def\ga{\alpha}
\def\gb{\beta}

\def\gd{\delta}
\def\ge{\epsilon}

\def\ggg{\gamma} % \gg = >>

\def\gk{\kappa}

\def\gq{\theta}

\def\gs{\sigma}

\def\gz{\zeta}

\def\gS{\Sigma}

\def\cB{{\mathcal B}}

\def\cD{{\mathcal D}}

\def\cF{{\mathcal F}}

\def\cM{{\mathcal M}}

\def\cW{{\mathcal W}}

\def\bfmath#1{\boldsymbol{#1}}

\def\bfk{{\bfmath{k}}}

\def\bfv{{\bfmath{v}}}

\def\bfgq{\bfmath{\gq}}
\def\A{p^{(x,z,s,t,\bfgq)}} % Lattice pmf notation
\def\cf#1#2{c_{#1,#2}^{(x,z,t,\bfgq)}} % Coefficients of f
\def\K{K} % Constant in Lemma 1
\def\Ca{C} % First constant in Proposition 1
\def\Cb{D} % Second constant in Proposition 1
\def\D{E} % First constant in Proposition 3
\def\E{F} % Constant in Proposition 4
\def\e{e} % Coefficient for bridge probabilities (Proposition 4)
\def\WF{\bbW\bbF}
\def\Cc{\widetilde{\Ca}} % Constant in proof of complexity proposition

\def\cdf{{\sc cdf}}
\def\pmf{{\sc pmf}}
\def\pdf{{\sc pdf}}
\def\sde{{\sc sde}}

\endlocaldefs

\begin{document}
\begin{frontmatter}
\title{Exact simulation of the Wright-Fisher diffusion}
\runtitle{Simulating the Wright-Fisher diffusion}
\author{\fnms{Paul A.}\ \snm{Jenkins}\corref{}\ead[label=e1]{p.jenkins@warwick.ac.uk}\thanksref{t1}}
\thankstext{t1}{Supported in part by EPSRC Research Grant EP/L018497/1.}
\address{Paul Jenkins\\Department of Statistics\\ ~\& Department of Computer Science\\ University of Warwick\\ Coventry CV4 7AL\\ United Kingdom\\ \printead{e1}}
\and
\author{\fnms{Dario} \snm{Span\`o}\ead[label=e2]{d.spano@warwick.ac.uk}}
\address{Dario Span\`o\\Department of Statistics\\ University of Warwick\\ Coventry CV4 7AL\\ United Kingdom\\ \printead{e2}}
\affiliation{University of Warwick}
\runauthor{P. A. Jenkins \& D. Span\`o}

\begin{abstract}
The Wright-Fisher family of diffusion processes is a widely used class of evolutionary models. However, simulation is difficult because there is no known closed-form formula for its transition function. In this article we demonstrate that it is in fact possible to simulate \emph{exactly} from a broad class of Wright-Fisher diffusion processes and their bridges. For those diffusions corresponding to reversible, neutral evolution, our key idea is to exploit an eigenfunction expansion of the transition function; this approach even applies to its infinite-dimensional analogue, the Fleming-Viot process. We then develop an exact rejection algorithm for processes with more general drift functions, including those modelling natural selection, using ideas from retrospective simulation. Our approach also yields methods for exact simulation of the moment dual of the Wright-Fisher diffusion, the ancestral process of an infinite-leaf Kingman coalescent tree. We believe our new perspective on diffusion simulation holds promise for other models admitting a transition eigenfunction expansion.
\end{abstract}


\end{frontmatter}

\section{Introduction}
Monte Carlo simulation of diffusion processes is of great interest, as it underlies methods of statistical inference from discrete observations in a variety of applications \citep[e.g.][]{gol:wil:2006, gol:wil:2008, chi:etal:ip, bla:sor:2014, bla:etal:2016}. Our interest in this paper is in the \emph{Wright-Fisher} diffusion. This process is widely used for inference, especially in genetics, where it serves  as a model for the evolution of the frequency $X_t \in [0,1]$ of a genetic variant, or \emph{allele}, in a large randomly mating population. If there are two alternative alleles then the diffusion obeys a one-dimensional stochastic differential equation (\sde{})
\begin{equation}
	\label{eq:WFSDE}
dX_t = \gamma(X_t)dt + \sqrt{X_t(1-X_t)}dB_t, \hspace{10pt} X_0 = x_0,\: t \in [0,T].
\end{equation}
The drift coefficient, $\gamma:[0,1]\to\bbR$, can encompass a variety of evolutionary forces. For example, $\gamma=\gb$ where
\begin{equation}
\label{eq:selection-drift}
\gb(x) = \frac{1}{2}[\theta_1(1-x) - \theta_2x] + \gs x(1-x)[x + h(1-2x)],
\end{equation}
describes a process with recurrent mutation between the two alleles, governed by parameters $\theta_1$ and $\theta_2$, and with (diploid) natural selection causing fitness differences between individuals with different numbers of copies of the allele, governed by parameters $\gs$ and $h$. There is much interest among geneticists in inference from this and related diffusions \citep[e.g.][]{wil:etal:2005, bol:etal:2008, gut:etal:2009, mal:etal:2012}, and in the characteristics of the trajectories themselves \citep{sch:etal:2013, zha:etal:2013}, as discretely observed data are becoming more and more available (for example, as genetic time series data for ancient human DNA and for viral evolution within hosts). Beyond genetics, Wright-Fisher diffusions have been considered for applications in several other fields. For example, in finance they have been used as models for time-evolving regime probability, discount coefficients or state price \citep[e.g.] []{del:shi:2002, Gourieroux2006475}; they have been proposed in biophysics as a model for ion channel dynamics \citep{dan:etal:2010, dan:etal:2012}; they have been studied as hidden Markov signals in filtering problems \citep{Chaleyat2009, PR14, PRS14}; and in Bayesian statistics they have been proposed as models for time-evolving priors  \citep{12348926, 6582111, gri:spa:2013, men:rug:2016}.

Simulation from equation \eqref{eq:WFSDE} is highly nontrivial because there is no known closed form expression for the transition function of the diffusion, even in the simple case $\ggg(x) \equiv 0$. In the absence of a method of exact simulation, it is necessary to turn to approximate alternatives such as an Euler-Maruyama scheme. Standard Euler-type methods fail here because simulated paths can easily leave the state space $[0,1]$, and moreover standard assumptions for weak and strong convergence typically require that the diffusion coefficient be Lipschitz continuous \citep[see][]{klo:pla:1999}. Consequently, a number of specialized time-discretization methods have been developed for the Wright-Fisher diffusion with various drifts; when the drift is of the form of $\gb(x)$, see \citet{sch:1996} for $\theta_1 = \theta_2 = \gs = 0$, \citet{dan:etal:2012} for $\gs = 0$, $\theta_1, \theta_2 > 0$, \citet{sch:etal:2013} for $\theta_1 = \theta_2 = 0$, $h = 1/2$; and \citet{neu:szp:2014} for $\gs = 0$, $\theta_1, \theta_2 \geq 1$. 
Other approaches include truncating a spectral expansion of the transition function \citep{luk:etal:2011, son:ste:2012, ste:etal:2013}, or numerical solutions of the Kolmogorov equations \citep{wil:etal:2005, bol:etal:2008, gut:etal:2009}. 
The error introduced by these methods can be difficult to quantify and must often be tested empirically.

In this paper we develop a novel and \emph{exact} method for simulating the Wright-Fisher diffusion with \emph{general} drift, as well as the corresponding bridges. By ``exact'' we mean that samples from the finite-dimensional distributions of the target diffusion can be recovered (up to the precision of a computer) without any approximation error. We build up our algorithm in stages, addressing how to simulate exactly from each of the following:
\begin{enumerate}
\item \label{item:neutral}The neutral Wright-Fisher diffusion; that is, with drift
\begin{equation}
\label{eq:neutral-drift}
\ga(x) = \frac{1}{2}[\theta_1(1-x) - \theta_2x],
\end{equation}
 where $\gq_1,\gq_2 > 0$ (\sref{sec:WF}).
\item \label{item:bridge} The neutral Wright-Fisher bridge (\sref{sec:WFbridge}); informally, this is the process
\[
(X_t)_{t\in[0,T]}\mid X_T = y.
\]
\item \label{item:nonneutral}The Wright-Fisher diffusion and its bridges with a very general class of drift functions (defined later), including drift $\gb(x)$ of the form \eqref{eq:selection-drift} when $\gq_1,\gq_2 > 0$ (\sref{sec:nonneutralWF}).
\end{enumerate}
To achieve step \ref{item:neutral} the key approach is to exploit an eigenfunction expansion representation of the transition function \citep[see][for review]{gri:spa:2010}. The expansion admits a probabilistic interpretation and therefore lends itself to simulation techniques, but these techniques are not straightforward to implement because the distributions involved are known only in infinite series form. Despite this hurdle, here we show that is possible to perform such simulation without error. The technique is very general and so we develop this section not just for the Wright-Fisher diffusion but for the \emph{Fleming-Viot process}, its infinite-dimensional generalization.

To achieve step \ref{item:bridge} we obtain a new probabilistic description of the transition function of a neutral Wright-Fisher \emph{bridge}. This is again complicated by the appearance of distributions known only in infinite series form, but from which (we show) realizations can still be obtained by evaluating only a finite number of terms in the series.

Finally, we generalize these techniques to nonneutral processes in step \ref{item:nonneutral}. The eigenfunction expansion for a nonneutral process \citep{bar:etal:2000} is probabilistically intractable, so we take a different approach: we use the simulated neutral processes as candidates in a rejection algorithm. 
This uses a retrospective approach similar to that of the ``exact algorithms'' of \citet{bes:rob:2005}, \citet{bes:etal:2006:B, bes:etal:2008}, and \citet{pol:etal:2016}, which can return exact samples from a class of diffusions using \emph{Brownian motion} as the candidate for rejection. We defer a detailed description to \sref{sec:EA}, but for now we note that a direct application of these techniques would require that the target diffusion satisfy a number of regularity conditions, the most stringent perhaps that its law be absolutely continuous with respect to Brownian motion. The Wright-Fisher diffusion \eqref{eq:WFSDE} fails in this regard, first because of its nonunit diffusion coefficient and second because of its finite boundaries. Although the first problem is easily solved via a Lamperti transformation [also known as Fisher's transformation when applied to \eqref{eq:WFSDE}], it is not clear how to deal with the second. \citet{bes:etal:2008} point out that their exact algorithm can be adopted to the case of two finite entrance boundaries, but this approach requires a further technical condition [(A3) below] which does not always hold here. When it does hold, this approach becomes arbitrarily inefficient when the diffusion is proximate to the boundary \citep{jen:arxiv}; in any case, the boundaries of the Wright-Fisher diffusion can be exit, regular reflecting, or entrance, depending on the parameters of $\gb(x)$. 
But now we are armed with the ability to simulate the neutral Wright-Fisher process, which serves as a far more promising candidate than Brownian motion in a rejection algorithm; specifically, it is known that the law of a nonneutral process is absolutely continuous with respect to its neutral counterpart \citep{daw:1978, eth:kur:1993}. We develop these ideas in full in \sref{sec:nonneutralWF}. We also remark that a related approach is taken by \citet{sch:etal:2013}, who were interested in simulating nonneutral Wright-Fisher bridges in the absence of mutation. In this context one can condition each sample path to remain in $(0,1)$ (otherwise the path could be absorbed at 0 or 1 and could not terminate at a pre-specified point), rendering the boundaries inaccessible. They show that the appropriate candidate in this case is a Bessel process of dimension 4, whose boundary at $0$ is also of entrance type. However, their method is not exact in the sense given above, since the rejection probabilities are approximated via numerical integration. Furthermore, the Radon-Nikod\'ym derivative of the Wright-Fisher process with respect to the Bessel(4) process is not bounded (another approach developed in \citet{jen:arxiv} for a single entrance boundary suffers a similar limitation). In any case a direct comparison with our method is not possible since here we tackle $\gq_1, \gq_2 > 0$ rather than $\gq_1 = \gq_2 = 0$.

The remainder of the paper is structured as follows. In Section \ref{sec:neutralperformance} we discuss practical considerations of the algorithm; \sref{sec:nonneutralbridge} fills in one last gap by showing how to construct a nonneutral Wright-Fisher bridge; Section \ref{sec:discussion} discusses extensions of the algorithm; and \sref{sec:proofs} contains the proofs of our results.
 
\section{Simulating the neutral Wright-Fisher process}
\label{sec:WF}
In this section we demonstrate how exact simulation from the neutral Wright-Fisher diffusion can be achieved. To aid the exposition we first focus on a one-dimensional process, and then later extend this idea to the Fleming-Viot process.
\subsection{A transition density expansion in one dimension}
Consider the diffusion satisfying \eqref{eq:WFSDE} with drift \eqref{eq:neutral-drift}. Denote its law by $\WF_{\ga,x}$ and its transition density $f(x,\cdotp;t)$. Throughout this paper we assume $\theta_1,\theta_2 > 0$; then $f(x,\cdotp;t)$ is a probability density. We will exploit the following probabilistic representation of the transition function's eigenfunction expansion \citep{gri:1979:AAP11:310, tav:1984, eth:gri:1993, gri:spa:2010}:
\begin{equation}
\label{eq:transition-dual}
f(x,y;t) = \sum_{m=0}^\infty q_m^\theta(t)\sum_{l=0}^m \cB_{m,x}(l)\cD_{\theta_1 + l, \theta_2 + m-l}(y),
\end{equation}
where
\[
\cB_{m,x}(l) = \binom{m}{l}x^l(1-x)^{m-l}, \qquad l=0,1,\ldots, m,
\]
is the probability mass function (\pmf{}) of a binomial random variable,
\[
\cD_{\theta_1, \theta_2}(y) = \frac{1}{B(\gq_1,\gq_2)}
y^{\theta_1 - 1}(1-y)^{\theta_2 - 1},
\]
is the probability density function (\pdf{}) of a beta random variable, $\theta = \theta_1 + \theta_2$, and $\{q_m^\theta(t) : m=0,1,\ldots \}$ are the transition functions of a certain death process $A^\gq_\infty(t)$ with an entrance boundary at $\infty$. More precisely, let $\{A^\theta_n(t): t\geq 0\}$ be a pure death process on $\bbN$ such that $A^\theta_n(0) = n$ almost surely and whose only transitions are $m \mapsto m-1$ at rate $m(m+\theta-1)/2$, for each $m=1,2,\ldots, n$. Then $q_m^\theta(t) = \lim_{n\to\infty}\bbP(A^\theta_n(t) = m)$.

The representation \eqref{eq:transition-dual} has a natural interpretation in terms of Kingman's coalescent, which is the moment dual to the Wright-Fisher diffusion. The ancestral process $A_\infty(t)$ represents the number of lineages surviving a time $t$ back in an infinite-leaf coalescent tree, when lineages are lost both by coalescence and by mutation. For our purposes, the mixture expression \eqref{eq:transition-dual} also provides an immediate method for \emph{simulating} from $f(x,\cdotp; t)$. We summarize this idea in Algorithm \ref{alg:f}, which first appeared in \citet{gri:li:1983}.

Steps \ref{f2} and \ref{f3} of Algorithm \ref{alg:f} are straightforward, but Step \ref{f1} requires the \pmf{} $\{q_m^\theta(t) : m=0,1,\ldots \}$, which is not available in closed form. \citet{gri:li:1983} used a numerical approximation, but in the following subsection we show it is in fact possible to simulate from this distribution without error.

\begin{algorithm}[t]
\DontPrintSemicolon
Simulate $A^\theta_\infty(t)$.
\nllabel{f1}\;
Given $A^\theta_\infty(t) = m$, simulate $L \sim \text{Binomial}(m,x)$.\nllabel{f2}\;
Given $L = l$, simulate $Y \sim \text{Beta}(\theta_1 + l, \theta_2 + m - l)$.\nllabel{f3}\;
\Return{$Y$}.\;
\caption{Simulating from the transition density $f(x,\cdotp; t)$ of the neutral Wright-Fisher diffusion with mutation.}\label{alg:f}
\end{algorithm}

\subsection{Simulating the ancestral process of Kingman's coalescent}
\label{sec:ancestral}
Our goal in this subsection is to obtain \emph{exact} samples from the discrete random variable with \pmf{} $\{q_m^\theta(t) : m=0,1,\ldots \}$. Were each $q_m^\theta(t)$ available in closed form, then standard inversion sampling would return exact samples from this distribution \citep[see for example][Ch.~2]{dev:1986}: for $U \sim \text{Uniform}[0,1]$,
\[
\inf\left\{M \in \bbN: \sum_{m=0}^M q_m^\theta(t) > U\right\}
\]
is distributed according to $\{q_m^\theta(t) : m=0,1,\ldots \}$. However, $q_m^\theta(t)$ is known only as an infinite series \citep{gri:1980, tav:1984}:
\begin{equation}
\label{eq:qm}
\begin{split}
q_m^\theta(t) &= \sum_{k=m}^\infty (-1)^{k-m}a_{km}^\theta e^{-k(k+\theta-1)t/2}, \text{ where}\\ a_{km}^\theta &= \frac{(\theta + 2k-1)(\theta + m)_{(k-1)}}{m!(k-m)!}.
\end{split}
\end{equation}
Here we have used the notation $a_{(x)} := \Gamma(a+x)/\Gamma(a)$ for $a > 0$ and $x \geq -1$.

Despite the apparently infinite amount of computation needed to evaluate \eqref{eq:qm}, we now show that it is nonetheless possible to return exact samples from this distribution by a variant of the \emph{alternating series method} \citep[Ch.~4]{dev:1986}, which we summarize for a discrete random variable $X$ on $\bbN$ as follows. Suppose $X$ has \pmf{} $\{p_m: m=0,1,\ldots\}$ of the form
\[
p_m = \sum_{k=0}^\infty (-1)^k b_k(m),
\]
and such that
\begin{equation}
\label{eq:monotone}
b_k(m) \downarrow 0 \text{ as }k \to \infty, \text{ for each }m.
\end{equation}
Then for each $M, K \in \bbN$,
\[
T^-_K(M) := \sum_{m=0}^M \sum_{k=0}^{2K+1} (-1)^k b_k(m) \leq \sum_{m=0}^M p_m \leq \sum_{m=0}^M \sum_{k=0}^{2K} (-1)^k b_k(m) =: T^+_K(M).
\]
Furthermore, $T^-_K(M) \uparrow \bbP(X \leq M)$ and $T^+_K(M) \downarrow \bbP(X \leq M)$ as $K \to \infty$. Hence, for $U \sim \text{Uniform}[0,1]$ and 
\[
K_0(M) := \inf\left\{K \in \bbN : T^-_K(M) > U \text{ or } T^+_K(M) < U\right\},
\]
the quantity $\inf\left\{M \in \bbN: T^-_{K_0(M)}(M) > U \right\}$ can be computed from finitely many terms and is exactly distributed according to \mbox{$\{p_m: m=0,1,\ldots\}$}.

This approach can be applied---with some modification---to $\{q_m^\theta(t) : m=0,1,\ldots \}$ by the following proposition, which says that the required condition \eqref{eq:monotone} holds with the possible exception of the first few terms in $m$. For those exceptional terms, \eqref{eq:monotone} still holds beyond the first few terms in $k$, and there is an easy way to check when this condition has been reached.

\begin{proposition}
\label{prop:qm}
Let $b^{(t,\theta)}_k(m) = a_{km}^\theta e^{-k(k+\theta-1)t/2}$, the relevant coefficient in \eqref{eq:qm}, and let
\begin{equation}
\label{eq:Cm}
\Ca^{(t,\gq)}_m := \inf\left\{ i \geq 0: b^{(t,\theta)}_{i+m+1}(m) < b^{(t,\theta)}_{i+m}(m)\right\}.
\end{equation}
Then
\begin{enumerate}[(i)]
\item $\Ca^{(t,\gq)}_m < \infty$, for all $m$. 
\item \label{item:monotonic} $b^{(t,\theta)}_k(m) \downarrow 0$ as $k \to \infty$ for all $k \geq m+\Ca^{(t,\gq)}_m$, and
\item $\Ca^{(t,\gq)}_m = 0$ for all $m > \Cb^{(t,\gq)}_0$, where for $\ge\in[0,1)$,
\end{enumerate}
\begin{equation}
\label{eq:C}
\ds \Cb^{(t,\gq)}_\ge := \inf\left\{ k \geq \left(\frac{1}{t} - \frac{\theta + 1}{2}\right)\vee 0: (\theta + 2k + 1)e^{-\frac{(2k+\theta)t}{2}} < 1-\ge\right\}.
\end{equation}
\end{proposition}
(The parameter $\ge$ is introduced for later use.) As a result of \propref{prop:qm}, we need only to make the following adjustment to the alternating series method: If $m \leq \Cb^{(t,\gq)}_0$ then precompute terms in $q^\theta_m(t)$ until the first time that the coefficients in \eqref{eq:qm} begin to decay; we know by \propref{prop:qm}(\ref{item:monotonic}) that this decay then continues indefinitely. To allow for the number of computed coefficients to depend on $m$ we introduce $\bfk = (k_0,k_1,\ldots,k_M)$ and 
\begin{equation}
\label{eq:T}
S^-_\bfk(M) := \sum_{m=0}^M \sum_{i=0}^{2k_m+1} (-1)^{i} b^{(t,\theta)}_{m+i}(m), \qquad S^+_\bfk(M) := \sum_{m=0}^M \sum_{i=0}^{2k_m} (-1)^{i} b^{(t,\theta)}_{m+i}(m).
\end{equation}
We summarize this procedure in Algorithm \ref{alg:q}.

Of course, the \emph{false} condition in line \ref{q15} of Algorithm \ref{alg:q} is never met, but \propref{prop:qm} guarantees that the algorithm still halts in finite time. Further performance considerations of this algorithm are discussed in \sref{sec:neutralperformance}. Also note that computed coefficients $a^\gq_{km}$ in \eqref{eq:qm} can also be stored for future calls to this algorithm.

\begin{algorithm}[t]
\DontPrintSemicolon
Set $m \longleftarrow 0$, $k_0 \longleftarrow 0$, $\bfk \longleftarrow (k_0)$.\;
Simulate $U \sim \text{Uniform}[0,1]$.\;
\Repeat{false\nllabel{q15}}{
Set $k_m \longleftarrow \lceil \Ca^{(t,\gq)}_m/2 \rceil$ [eq.~\eqref{eq:Cm}]. \nllabel{qprecomputation}\;
\While{$S_\bfk^-(m) < U < S_\bfk^+(m)$}{Set $\bfk \longleftarrow \bfk + (1,1,\ldots,1)$\nllabel{q6}}\;
\uIf{$S_\bfk^-(m) > U$}{\Return{$m$}}
\ElseIf{$S_\bfk^+(m) < U$}{Set $\bfk \longleftarrow (k_0,k_1,\ldots,k_m, 0)$.\;
Set $m \longleftarrow m+1$.}
}
\caption{Simulating from the ancestral process $A_\infty(t)$ of Kingman's coalescent with mutation.}\label{alg:q}
\end{algorithm}

\subsection{A transition density expansion in higher dimensions}
\label{sec:multidimensionalWF}
It is worth pointing out that an interesting by-product of \propref{prop:qm} (and of Algorithm \ref{alg:q}) is the possibility of simulating exactly from the transition function of the (parent-independent) neutral Wright-Fisher diffusion \emph{in any dimension}, even in infinite dimensions. 
Wright-Fisher diffusions in $d$ dimensions can be seen as $d$-dimensional projections of a so-called neutral (parent-independent) Fleming-Viot measure-valued diffusion $\mu=(\mu_t:t\geq 0)$ with state space ${\cal M}_1(E)$, the set of all the probability measures on a given (Polish) type space $E$, equipped with the Borel sigma-algebra induced by the weak convergence topology. Given a total mutation parameter $\theta$ and a mutation distribution $P_0\in{\cal M}_1(E)$, the process $\mu$ is reversible with stationary distribution given by the Dirichlet process with parameter $(\theta, P_0)$, here denoted with $\Pi_{\theta, P_0}$, characterized by Dirichlet finite-dimensional distributions:
$$\Pi_{\theta, P_0}\left(\bigcap_{i=1}^d\{\mu(A_i)\in dx_i,\}\right)\propto \left[\prod_{i=1}^d x_i^{\theta P_0(A_i)-1} dx_i\right]\mathbb{I}_{\Delta_{(d-1)}}(x_1,\ldots,x_d)$$
for any $d$ and every measurable partition $A_1,\ldots,A_d$ of $E$,
where $\Delta_{(d-1)}=\{(x_1,\ldots,x_d)\in[0,1]^d:\sum_1^dx_i=1\}.$

The  transition function describing the evolution of $\mu$ admits a probabilistic series expansion as mixture of (posterior) Dirichlet processes:
\begin{multline}
p(\mu, d\nu;t)=
\sum_{m=0}^\infty q^\theta_m(t)\int_{E^m}\mu^{\otimes m}(d\xi_1,\ldots,d\xi_m)\Pi_{\theta+m, \frac{m}{\theta+m}\eta_m+\frac{\theta}{\theta+m}P_0}(d\nu),\\
t\geq 0, \mu,\nu\in{\cal M}_1(E),
\label{eq:fvtf}
\end{multline}
where $\mu^{\otimes n}$ denotes the $n$-fold $\mu$-product measure and $\eta_m:=m^{-1}\sum_{i=1}^m\delta_{\xi_i}$ \citep[see][]{eth:gri:1993}.
The coefficients of the series expansion are given by \emph{i.i.d.}\ samples (the $\xi$-random variables) from the starting measure, $\mu$, of random size given by the coalescent lines-of-descent counting process $A^\theta_\infty(t)$ with distribution $q_m^\theta(t)$. An algorithm for simulating from the transition function \eqref{eq:fvtf} is thus the following modification of Algorithm \ref{alg:f}.
\\

 \begin{algorithm}[H]
\DontPrintSemicolon
Simulate $A^\theta_\infty(t)$.\nllabel{fv1}\;
Given $A^\theta_\infty(t) = m$, simulate $\xi_1,\ldots,\xi_m \overset{iid}{\sim} \mu$.\nllabel{fv2}\;
Given $m^{-1}\sum_{i=1}^m\delta_{\xi_i} = \eta_m$, simulate 
$\nu \sim \Pi_{\theta+m, \frac{m}{\theta+m}\eta_n+\frac{\theta}{\theta+m}P_0}$.\nllabel{fv3}\;
\Return{$\nu$}.\;
\caption{Simulating from the transition density $p(\mu, \cdotp;t)$ of the neutral Fleming-Viot process with parent-independent mutation.}\label{alg:fv}
\end{algorithm}
~\\

Notice that step \ref{fv3} requires sampling a (potentially infinite-dimensional) random measure distributed according to a Dirichlet process. Techniques for exact sampling from a Dirichlet process have been available in the literature [e.g.\ \citet{pap:rob:2008} and \citet{wal:2007}] for quite some time. Hence Algorithm \ref{alg:q} provides a way of filling the only missing gap (Step \ref{fv1} of Algorithm \ref{alg:fv}) to make the transition function \eqref{eq:fvtf} viable for exact simulation.
When $E$ consists of $d$ points ($d\in\mathbb N$) the process reduces to the $(d-1)$-dimensional Wright-Fisher diffusion, thus Algorithm \ref{alg:f} is viable for exact simulation of neutral $(d-1)$-dimensional extensions of the Wright-Fisher diffusion \eqref{eq:WFSDE} with drift \eqref{eq:neutral-drift}.

\section{Simulating a neutral Wright-Fisher bridge}
\label{sec:WFbridge}
In this section we demonstrate how exact simulation from the neutral Wright-Fisher diffusion \emph{bridges} can be achieved, via a new probabilistic description of its transition density. For the remainder of the paper we return to processes of dimension one.
\subsection{A transition density expansion}
\sref{sec:WF} provides a method for returning exact samples from $f(x,\cdotp;t)$ for any fixed $x \in [0,1]$ and $t > 0$. 
The density of a point $y \in (0,1)$ at time $s$ in a Wright-Fisher bridge from $x$ at time $0$ to $z$ at time $t$ is given by \citep{fit:etal:1992, sch:etal:2013}:
\begin{equation}
\label{eq:bridge-transition}
f_{z,t}(x,y; s) = \frac{f(x,y; s)f(y,z;t-s)}{f(x,z; t)}, \qquad  0 < s < t,
\end{equation}
with $f(\cdotp,\cdotp; \cdotp)$ as in \eqref{eq:transition-dual}. Motivated by \eqref{eq:transition-dual}, our goal is to facilitate easy simulation from $f_{z,t}(x,y; s)$ by putting it into mixture form. For the rest of this section we assume $0 < x,y,z < 1$.

\begin{proposition}
\label{prop:WFbridge}
The transition density of a Wright-Fisher bridge has expansion
\begin{equation}
\label{eq:bridge-mixture}
f_{z,t}(x,y; s) = \sum_{m=0}^\infty \sum_{k=0}^\infty \sum_{l=0}^m \sum_{j=0}^k \A_{m,k,l,j} \cD_{\theta_1 + l + j, \theta_2 + m + k - l - j}(y),
\end{equation}
where 
\begin{align}
\label{eq:pmfN4}
\A_{m,k,l,j} &= \cB_{m,x}(l)\cD_{\gq_1 + j, \gq_2 + k - j}(z)\cD\cM_{\gq_1 + l, \gq_2 + m-l;k}(j)\frac{q^\theta_m(s)q^\theta_k(t-s)}{f(x,z;t)},
\end{align}
for $0 \leq l \leq m$ and $0 \leq j \leq k$, and $\A_{m,k,l,j} =0$ otherwise, where
\[
\cD\cM_{a,b;k}(j) = \binom{k}{j}\frac{B(a+j,b+k-j)}{B(a,b)}
\]
is the \pmf{} of a beta-binomial random variable on $\{0,1,\dots, k\}$.
\end{proposition}

By \propref{prop:WFbridge}, we recognize equation \eqref{eq:bridge-mixture} as a mixture of beta-distributed random variables, with mixture weights defining a \pmf{} $\{\A_{m,k,l,j}: m,k,l,j \in \bbN\}$ on $\bbN^4$. Thus, the following algorithm returns exact samples from $f_{z,t}(x,y; s)$.
\\

\begin{algorithm}[H]
\DontPrintSemicolon
Simulate $(M,K,L,J) \sim \{\A_{m,k,l,j}: m,k,l,j \in \bbN\}$ [eq.~\eqref{eq:pmfN4}].\nllabel{fbridge1}\;
Given $(M,K,L,J) = (m,k,l,j)$, simulate $Y \sim \text{Beta}(\theta_1 + l + j, \theta_2 + m + k - l - j)$.\nllabel{fbridge2}\;
\Return{$Y$}
\caption{Simulating from the transition density $f_{z,t}(x,y; s)$ of a bridge of the neutral Wright-Fisher diffusion with mutation.}
\label{alg:fbridge}
\end{algorithm}
~\\

Again, while step \ref{fbridge2} of Algorithm \ref{alg:fbridge} is straightforward, Step \ref{fbridge1} is complicated by the appearance of $q_m^\theta(s)q^\theta_k(t-s)/f(x,z;t)$ in \eqref{eq:pmfN4}; each term in this ratio is known only as an infinite series, as we have seen. We address this in the following subsection.

\subsection{Simulating from the discrete random variable on $\bbN^4$ with \pmf{}\\ \mbox{$\{\A_{m,k,l,j}: m,k,l,j \in \bbN\}$}}
We will apply the alternating series approach of \sref{sec:ancestral} separately to each of $q_m^\theta(s)$, $q^\theta_k(t-s)$, and $f(x,z;t)$, and then combine these to obtain monotonically converging upper and lower bounds on \eqref{eq:pmfN4}. The first two terms have been dealt with already in \sref{sec:ancestral}, so it remains to take a similar approach for $f(x,z;t)$. Note that this problem---the pointwise evaluation of $f(x,z;t)$---is separate from (and in this case, harder than) actually simulating from $f(x,\cdotp;t)$.

To employ the alternating series approach, use \eqref{eq:transition-dual} and \eqref{eq:qm} to write
\begin{align}
f(x,z;t) &= \sum_{m=0}^\infty \sum_{k=m}^\infty (-1)^{k-m}\cf{k}{m}, \label{eq:transition2}\\
\text{where} \quad \cf{k}{m} &= a_{km}^\gq e^{-k(k+\gq-1) t/2} \bbE[\cD_{\gq_1 + L_m, \gq_2 + m - L_m}(z)], \notag
\end{align}
and $L_m \sim \text{Binomial}(m,x)$. Our strategy is to group the triangular array of coefficients $(\cf{k}{m})_{k \geq m}$ in such a way that, with the exception of the first few terms, they exhibit a property analogous to \eqref{eq:monotone}. We will compare terms in the sequence $(d_i)_{i=0,1,\ldots}$ of antidiagonals, defined by
\begin{align}
\label{eq:antidiagonal}
d_{2m} &=  \sum_{j=0}^{m} \cf{m+j}{m-j}, &
d_{2m+1} &= \sum_{j=0}^{m} \cf{m+1+j}{m-j}, & m &= 0,1,\ldots,
\end{align}
(see \fref{fig:dm}), and dropping the superscript for convenience. Notice that the coefficients within each entry of this sequence are all multiplied by the same sign in \eqref{eq:transition2}, so that $f(x,z;t) = d_0 - d_1 + d_2 - d_3 + \ldots$ will be our alternating sequence. The main complication in this approach is to find \emph{explicitly} the first $i$ for which the coefficients $(d_i)$ begin to decrease in magnitude. To this end we define
\begin{equation}
\label{eq:antidiagonal-decaying}
\D^{(t,\gq)} := \inf\left\{ m \geq 0: 2j \geq \Ca^{(t,\gq)}_{m-j} \text{ for all }j=0,\ldots,m\right\},
\end{equation}
which simply provides the first entry in $(d_{2m})$ for which every member of the corresponding antidiagonal is decaying as a function of its first index. We now need the following lemma.
\begin{lemma}
\label{lem:beta}
Let $L_m \sim \text{Binomial}(m,x)$ and
\begin{equation}
\label{eq:K}
\K^{(x,z)} := \left(\frac{\theta}{\theta_2}(1-z) \vee \frac{1+\theta}{1-z}\right)(1-x) + \left(1+\frac{1}{z}\right)\left[\frac{z\theta + 1}{\theta_1} \vee 1\right]x.
\end{equation}
Then for all $m \in \bbN$, 
\begin{equation}
\label{eq:beta}
\bbE[\cD_{\gq_1 + L_{m+1}, \gq_2 + m+1 - L_{m+1}}(z)] < \K^{(x,z)}\bbE[\cD_{\gq_1 + L_{m}, \gq_2 + m - L_{m}}(z)].
\end{equation}
\end{lemma}
Using \lemmaref{lem:beta} we are in a position to obtain the required analogue of property \eqref{eq:monotone}.
\begin{proposition}
\label{prop:lattice}
Let $\Cb^{(t,\gq)}_\ge$, $(d_i)_{i=0,1,\ldots}$, $\D^{(t,\gq)}$, and $\K^{(x,z)}$ be defined as in \eqref{eq:C}, \eqref{eq:antidiagonal}, \eqref{eq:antidiagonal-decaying}, and \eqref{eq:K}, respectively, and $\ge \in (0,1)$. Then
\[
d_{2m+2} < d_{2m+1} < d_{2m}
\]
for all $m \geq \D^{(t,\gq)} \vee \Cb^{(t,\gq)}_\ge \vee 2\K^{(x,z)}/\ge$.
\end{proposition}

We can now combine Propositions \ref{prop:qm} and \ref{prop:lattice} in order to construct a sequence amenable to simulation from $\{\A_{m,k,l,j}: m,k,l,j \in \bbN\}$ [eq.~\eqref{eq:pmfN4}], via the alternating series method.

\begin{proposition}
\label{prop:bes:etal:2008:1}
Define
\begin{equation}
\label{eq:E}
\E^{(s,t,\gq,x,z)}_{m,k,l,j} := \Ca^{(s,\gq)}_m \vee \Ca^{(t-s,\theta)}_k \vee \D^{(t,\gq)} \vee \Cb^{(t,\gq)}_\ge \vee 2\K^{(x,z)}/\ge,
\end{equation}
and
\begin{multline*}
\e_{m,k,l,j}(v) := \cB_{m,x}(l)\cD_{\gq_1 + j, \gq_2 + k - j}(z)\cD\cM_{\gq_1 + l, \gq_2 + m-l;k}(j)\\
\times \left. {\left[\ds\sum_{i=0}^{v} (-1)^i b_{m+i}^{(s,\gq)}(m)\right]\left[\ds\sum_{i=0}^{v} (-1)^i b_{k+i}^{(t-s,\gq)}(k) \right]}\middle/ {\ds\sum_{i=0}^{v+1} (-1)^i d_{i}} \right..
\end{multline*}
Then for $2v \geq \E^{(s,t,\gq,x,z)}_{m,k,l,j}$,
\[
\e_{m,k,l,j}(2v+1) < \e_{m,k,l,j}(2v+3) < \A_{m,k,l,j} < \e_{m,k,l,j}(2v+2) < \e_{m,k,l,j}(2v).
\]
\end{proposition}
In other words, for sufficiently large $v$ the odd and even terms in the sequence $(\e_{m,k,l,j}(v))_{v=0}^\infty$ provide monotonically converging lower and upper bounds on $\A_{m,k,l,j}$, respectively, and ``sufficiently large'' can be verified explicitly.

The above results are summarized in Algorithm \ref{alg:pmfN4}. To explore $\bbN^4$ we introduce for convenience a bijective pairing function $\gS : \bbN \to \bbN^4$, such that $\gS(n) = (m,k,l,j)$. As in Algorithm \ref{alg:q}, we also introduce $\bfv = (v_0,v_1,\ldots,v_N)$ and 
\begin{equation*}
\label{eq:V}
V^-_\bfv(N) := \sum_{n=0}^N \e_{\gS(n)}(2v_n+1), \qquad V^+_\bfv(N) := \sum_{n=0}^N \e_{\gS(n)}(2v_n).
\end{equation*}

\begin{algorithm}[H]
\DontPrintSemicolon
Set $n \longleftarrow 0$, $v_0 \longleftarrow 0$, $\bfv \longleftarrow (v_0)$.\;
Simulate $U \sim \text{Uniform}[0,1]$.\;
\Repeat{false}{
Set $v_n \longleftarrow \lceil \E_{\gS(n)}/2\rceil$ [eq.~\eqref{eq:E}]. \nllabel{pmfN4precomputation}\;
\While{$V_\bfv^-(n) < U < V_\bfv^+(n)$}{Set $\bfv \longleftarrow \bfv + (1,1,\ldots,1)$}\;
\uIf{$V_\bfv^-(n) > U$}{\Return{$\gS(n)$}}
\ElseIf{$V_\bfv^+(n) < U$}{Set $\bfv \longleftarrow (v_0,v_1,\ldots,v_n, 0)$.\;
Set $n \longleftarrow n+1$.}
}
\caption{Simulating from the discrete random variable on $\bbN^4$ with \pmf{} \mbox{$\{\A_{m,k,l,j}: m,k,l,j \in \bbN\}$}.}\label{alg:pmfN4}
\end{algorithm}

\section{Performance of algorithms for neutral processes}
\label{sec:neutralperformance}
There are several easy improvements to the underlying Algorithm \ref{alg:q}. For example, we are free to vary the order of inspection of each $m$ by any finite permutation of $\bbN$, and we found a dramatic improvement by radiating outwards from (an approximation of) the mode of $\{q_m^\theta(t): m=0,1,\ldots\}$ than to start at $m=0$ and work upwards. Our approximation used $\mu^{(t,\gq)}$ in \thmref{thm:gri:1984} below. It may also be possible to improve on Algorithm \ref{alg:q} by allowing different $q_m^\theta(t)$ to be refined at different rates, i.e.\ by using a vector other than $(1,1,\dots,1)$ in Step \ref{q6}; we do not explore that here.

A crucial quantity governing the efficiency of our algorithms is the number of coefficients $b^{(t,\theta)}_k(m)$ we must compute in Algorithm \ref{alg:q}; these in turn depend on the quantities $\Cb^{(t,\gq)}_0$ and $\Ca^{(t,\gq)}_m$ (recall \propref{prop:qm}). 
These quantities are in general manageably small, suggesting that the number of coefficients that need to be computed in Algorithms \ref{alg:q} and \ref{alg:pmfN4} (line \ref{qprecomputation} in each) should not be excessive. 

One exception to this observation is when $t$ is very small, for which the number of relevant coefficients grows quickly. The following result makes precise the complexity of Algorithm \ref{alg:q}.

\begin{proposition}
As $t\to 0$,
\begin{enumerate}[(i)]
\item $\Ca_m^{(t,\gq)} = O(t^{-1})$.
\item $\max_m \Ca_m^{(t,\gq)} = O(t^{-1}\log (t^{-1}))$.
\item $\Cb_0^{(t,\gq)} = o(t^{-(1+\gk)})$, for any $\gk > 0$.
\end{enumerate}
Let $N^{(t,\gq)}$ denote the total number of coefficients that must be computed in an implementation of Algorithm \ref{alg:q}. Then $\bbE[N^{(t,\gq)}] < \infty$, and in particular
\begin{enumerate}[(i)]
\item[(iv)] $\bbE[N^{(t,\gq)}] = o(t^{-(1+\gk)})$, for any $\gk > 0$.
\end{enumerate}
\label{prop:neutralcomplexity}
\end{proposition}

The growth in Algorithm \ref{alg:q} as $t\to 0$ is closely related to the well known numerical instability of \eqref{eq:qm} as $t \to 0$ \citep{gri:1984}, which afflicts any method based on the expansion \eqref{eq:transition-dual} \citep[or an expansion using a basis of orthogonal polynomials, which is equivalent to \eqref{eq:qm};][]{gri:spa:2010}. 
In any practical implementation of our algorithm we are obliged to use an approximation should the separation between two points be very small, and we found Algorithm \ref{alg:q} to fail for $t < 0.05$ or so. One option is to revert to the Euler-Maruyama approximation for small $t$. Alternatively, there has been much previous work in coalescent theory on approximating the distribution \eqref{eq:qm} \citep[e.g.][]{gri:1984, gri:2006, jew:ros:2014}; by inserting those approximations into Algorithm \ref{alg:q} they readily define new algorithms for approximate simulation of the dual \emph{diffusion}. We consider the following result, due to \citet[Theorem 4]{gri:1984}.
\begin{theorem}[\citet{gri:1984}]
\label{thm:gri:1984}
Suppose $\beta = \frac{1}{2}(\theta - 1)t$,  
and let 
\begin{align*}
\mu^{(t,\gq)} &= \frac{2\eta}{t}, & (\sigma^{(t,\gq)})^2 &= \begin{cases} \frac{2\eta}{t}(\eta + \beta)^2\left(1 + \frac{\eta}{\eta+\beta} - 2\eta\right)\beta^{-2}, & \beta \neq 0,\\
\frac{2}{3t}, & \beta = 0,
\end{cases} & \text{where}\\ && \eta &= \begin{cases} \frac{\beta}{e^\beta - 1}, & \beta \neq 0,\\
1, & \beta = 0.
\end{cases} 
\end{align*}
Then $\bbP\left((A_\infty^\gq(t) - \mu^{(t,\gq)})(\gs^{(t,\gq)})^{-1} \leq x\right) \to \Phi(x)$ as $t\to 0$, where $\Phi(\cdot)$ is the cumulative distribution function (\cdf{}) of a standard normal random variable.
\end{theorem}
[The statement in \citet{gri:1984} is missing the factor $\beta^{-2}$.] To apply this approximation in practice when $t$ is small, we replace line \ref{f1} in Algorithm \ref{alg:f} with
\begin{itemize}
\item[{\scriptsize \bf 1'}] Simulate $A_\infty(t) \sim N(\mu^{(t,\gq)},(\gs^{(t,\gq)})^2)$ and round it to the nearest nonnegative integer. \label{page:1'}
\end{itemize}
\subsection{Comparison with Euler-Maruyama simulation}
To check the correctness of our algorithm and to compare its performance to Euler-Maruyama simulation, we performed the following experiment. We fixed $\gq_1 = \gq_2 = 1/2$, explored various fixed values of $x$ and $t$, and for each parameter combination drew 10,000 samples from $f(x,\cdot; t)$ using Algorithm \ref{alg:f}. To quantify whether the resulting sample was consistent with the true distribution $f(x,\cdot; t)$, we performed a one-sample Kolmogorov-Smirnov (K-S) test. We then performed the same experiment instead using Euler simulation with various stepsizes $\gd$. For this purpose we used the Balanced Implicit Split Step (BISS) algorithm of \citet{dan:etal:2012}, an Euler-type algorithm with some advanced modifications that guarantee each sample path stays within $[0,1]$, and which is state-of-the-art for $\gq_1,\gq_2 > 0$. (Their algorithm has an additional tuning parameter $\ge$; we followed their recommendation and set $\ge = \gd/4$.)

To obtain an accurate expression for the \cdf{} of the true distribution for use within the K-S statistic, we exploited the fact that, in the special case $\gq_1 = \gq_2 = 1/2$, a Lamperti transformation of \eqref{eq:WFSDE} (or conversion to Stratonovich form) leads to
\[
X_t = \frac{1}{2}(1-\cos B_t),
\]
where $(B_t)_{t\geq 0}$ is a Brownian motion commenced from $\arccos(1-2x)$ and reflecting at 0 and $\pi$. A rapidly converging series expression for the \cdf{} of $B_t$ (and hence of $X_t$) is available \citep[eq.~(26)]{lin:2005}; the first 1000 terms in the series suggested convergence to machine precision and were used as the reference \cdf{}.

\begin{table}[p]
\caption{\label{tab:neutralWFresults}Comparison of exact simulation methods for the neutral Wright-Fisher diffusion. {\bf BISS}: Algorithm of \citet{dan:etal:2012} with stepsize $\gd$. {\bf Exact}: Algorithm \ref{alg:f}. {\bf Exact'}: Algorithm \ref{alg:f} with the approximation of \citet{gri:1984} described on p\pageref{page:1'}. Tabulated are the computing time needed to simulate 10,000 sample paths and the $p$-value of a K-S test applied to the resulting collection of endpoints. Paths are initiated at $X_0 = x$ and run for length $t$. Mutation parameters are $\gq_1 = \gq_2 = 1/2$.} 
\vspace{-10pt}
\begin{center}
{\scriptsize
\begin{tabular}{cc}
\multicolumn{2}{c}{$t=0.01$}\\\hline
\begin{tabular}{c c r@{.}l c }
\multicolumn{5}{c}{$x = 0.01$}\\\hline
Method & $\gd$ & \multicolumn{2}{c}{Time (s)} & $p$-value\\
\hline
\multirow{5}{*}{BISS} 	%& $10^{-2}$ 	& 0&02 	& $<10^{-100}$ 	\\%%%%%%%%%%%%%%%%%%%%%%%%%%%%%%%%%%%%%%%%%%%%%%%%%%%%%%%%%%%%%%
& $10^{-3}$ 	& 0&03 	& $<10^{-100}$ 	\\%%%%%%%%%%%%%%%%%%%%%%%%%%%%%%%%%%%%%%%%%%%%%%%%%%%%%%%%%%%%%%
& $10^{-4}$ 	& 0&05 	& $<10^{-100}$ 	\\%%%%%%%%%%%%%%%%%%%%%%%%%%%%%%%%%%%%%%%%%%%%%%%%%%%%%%%
& $10^{-5}$ 	& 0&30 	& $<10^{-100}$ 	\\%%%%%%%%%%%%%%%%%%%%%%%%%%%%%%%%%%%%%%%%%%%%%%%%%%%%%%%
	& $10^{-6}$ 	& 2&83 	& $<10^{-100}$ 	\\%%%%%%%%%%%%%%%%%%%%%%%%%%%%%%%%%%%%%%%%%%%%%%%%%%%%%%%
	& $10^{-7}$ 	& 27&61 	& $<10^{-100}$ 	\\%%%%%%%%%%%%%%%%%%%%%%%%%%%%%%%%%%%%%%%%%%%%%%%%%%%%%%%
%	& $10^{-8}$ 	& 276&21 	& $1.3\times 10^{-98}$ 	\\%%%%%%%%%%%%%%%%%%%%%%%%%%%%%%%%%%%%%%%%%%%%%%%%%%%%%%%
Exact' & -- 		& 0&19 	& 0.94%%%%%%%%%%%%%%%%%%%%%%%%%%%%%%%%%%%%%%%%%%%%%%%%%%%%%%%%%%
\end{tabular} &
\begin{tabular}{c c r@{.}l c }
\multicolumn{5}{c}{$x = 0.5$}\\\hline
Method & $\gd$ & \multicolumn{2}{c}{Time (s)} & $p$-value\\\hline
\multirow{5}{*}{BISS} %& $10^{-2}$ 	& 0&02 	& $6.4\times 10^{-11}$ 	\\%%%%%%%%%%%%%%%%%%%%%%%%%%%%%%%%%%%%%%%%%%%%%%%%%%%%%%%%%
  & $10^{-3}$ 	& 0&02 	& $8.0\times 10^{-3}$ 	\\%%%%%%%%%%%%%%%%%%%%%%%%%%%%%%%%%%%%%%%%%%%%%%%%%%%%%%%%%%%
 & $10^{-4}$ 	& 0&05 	& 0.36 	\\%%%%%%%%%%%%%%%%%%%%%%%%%%%%%%%%%%%%%%%%%%%%%%%%%%%%%%%%%%%%
  & $10^{-5}$ 	& 0&30 	& 0.22 	\\%%%%%%%%%%%%%%%%%%%%%%%%%%%%%%%%%%%%%%%%%%%%%%%%%%%%%%%%%%%%
  & $10^{-6}$ 	& 2&78 	& 0.30 	\\%%%%%%%%%%%%%%%%%%%%%%%%%%%%%%%%%%%%%%%%%%%%%%%%%%%%%%%%%%%%%
  & $10^{-7}$ 	& 27&44 	& 0.78 	\\%%%%%%%%%%%%%%%%%%%%%%%%%%%%%%%%%%%%%%%%%%%%%%%%%%%%%%%%%%%%%
%  & $10^{-8}$ 	& 274&25 	& 0.37 	\\%%%%%%%%%%%%%%%%%%%%%%%%%%%%%%%%%%%%%%%%%%%%%%%%%%%%%%%%%%%%%
Exact'  & -- 		& 0&17 	& 0.18%%%%%%%%%%%%%%%%%%%%%%%%%%%%%%%%%%%%%%%%%%%%%%%%%%%%%%%%%%%
\end{tabular} \\ \hline\\
\multicolumn{2}{c}{$t=0.05$}\\\hline
\begin{tabular}{c c r@{.}l c }
\multicolumn{5}{c}{$x = 0.01$}\\\hline
Method & $\gd$ & \multicolumn{2}{c}{Time (s)} & $p$-value\\\hline
\multirow{5}{*}{BISS} 	%& $10^{-2}$ 	& 0&02 	& $<10^{-100}$ \\%%%%%%%%%%%%%%%%%%%%%%%%%%%%%%%%%%%%%%%%%%%%%%%%%%%%%%%%%%%%%%
  & $10^{-3}$ 	& 0&04 	& $<10^{-100}$ \\%%%%%%%%%%%%%%%%%%%%%%%%%%%%%%%%%%%%%%%%%%%%%%%%%%%%%%%%%%%%%%
 & $10^{-4}$ 	& 0&16 	& $<10^{-100}$ \\%%%%%%%%%%%%%%%%%%%%%%%%%%%%%%%%%%%%%%%%%%%%%%%%%%%%%%%%%%%%%%
 & $10^{-5}$ 	& 1&40 	& $<10^{-100}$ \\%%%%%%%%%%%%%%%%%%%%%%%%%%%%%%%%%%%%%%%%%%%%%%%%%%%%%%%%%%%%%%
 & $10^{-6}$ 	& 13&77 	& $<10^{-100}$ \\%%%%%%%%%%%%%%%%%%%%%%%%%%%%%%%%%%%%%%%%%%%%%%%%%%%%%%%%%%%%%
 & $10^{-7}$ 	& 138&08 	& $<10^{-100}$ \\%%%%%%%%%%%%%%%%%%%%%%%%%%%%%%%%%%%%%%%%%%%%%%%%%%%%%%%%%%%%%
% & $10^{-8}$ 	& 1377&95 	& $<10^{-100}$ \\%%%%%%%%%%%%%%%%%%%%%%%%%%%%%%%%%%%%%%%%%%%%%%%%%%%%%%%%%%%%%
Exact & -- 		& 0&35 	& 0.09\\%%%%%%%%%%%%%%%%%%%%%%%%%%%%%%%%%%%%%%%%%%%%%%%%%%%%%%%%%
Exact' & -- 		& 0&17 	& 0.30%%%%%%%%%%%%%%%%%%%%%%%%%%%%%%%%%%%%%%%%%%%%%%%%%%%%%%%%%%%
\end{tabular} &
\begin{tabular}{c c r@{.}l c }
\multicolumn{5}{c}{$x = 0.5$}\\\hline
Method & $\gd$ & \multicolumn{2}{c}{Time (s)} & $p$-value\\\hline
\multirow{5}{*}{BISS}   %& $10^{-2}$ 	& 0&02 	& $1.6 \times10^{-15}$ 	\\%%%%%%%%%%%%%%%%%%%%%%%%%%%%%%%%%%%%%%%%%%%%%%%%%%%%%%%%%
 & $10^{-3}$ 	& 0&04 	& $9.0\times10^{-4}$ 	\\%%%%%%%%%%%%%%%%%%%%%%%%%%%%%%%%%%%%%%%%%%%%%%%%%%%%%%%%%%
& $10^{-4}$ 	& 0&16 	& 0.35 	\\%%%%%%%%%%%%%%%%%%%%%%%%%%%%%%%%%%%%%%%%%%%%%%%%%%%%%%%%%%%%
 & $10^{-5}$ 	& 1&39 	& 0.53 	\\%%%%%%%%%%%%%%%%%%%%%%%%%%%%%%%%%%%%%%%%%%%%%%%%%%%%%%%%%%%%
 & $10^{-6}$ 	& 13&66 	& 0.02 	\\%%%%%%%%%%%%%%%%%%%%%%%%%%%%%%%%%%%%%%%%%%%%%%%%%%%%%%%%%%
 & $10^{-7}$ 	& 137&06 	& 0.72 	\\%%%%%%%%%%%%%%%%%%%%%%%%%%%%%%%%%%%%%%%%%%%%%%%%%%%%%%%%%%
% & $10^{-8}$ 	& 1370&35 	& 0.68 	\\%%%%%%%%%%%%%%%%%%%%%%%%%%%%%%%%%%%%%%%%%%%%%%%%%%%%%%%%%%
Exact  & -- 		& 0&34 	& 0.64\\%%%%%%%%%%%%%%%%%%%%%%%%%%%%%%%%%%%%%%%%%%%%%%%%%%%%%%%%%%%
%& 1e-06 	& 13.78 	& 0.039 	\\%%%%%%%%%%%%%%%%%%%%%%%%%%%%%%%%%%%%%%%%%%%%%%%%%%%%%%%%%%
Exact' & -- 		& 0&17 	& 0.97%%%%%%%%%%%%%%%%%%%%%%%%%%%%%%%%%%%%%%%%%%%%%%%%%%%%%%%%%%%
\end{tabular} \\ \hline\\

\multicolumn{2}{c}{$t=0.5$}\\\hline
\begin{tabular}{c c r@{.}l c }
\multicolumn{5}{c}{$x = 0.01$}\\\hline
Method & $\gd$ & \multicolumn{2}{c}{Time (s)} & $p$-value\\
\hline
\multirow{5}{*}{BISS}%& $10^{-2}$ 	& 0&04 	& $<10^{-100}$ 	\\%%%%%%%%%%%%%%%%%%%%%%%%%%%%%%%%%%%%%%%%%%%%%%%%%%%%%%%%%%%%%%%
 	& $10^{-3}$ 	& 0&16 	& $<10^{-100}$ \\%%%%%%%%%%%%%%%%%%%%%%%%%%%%%%%%%%%%%%%%%%%%%%%%%%%%%%%%%%%%%%%
 	& $10^{-4}$ 	& 1&43 	& $<10^{-100}$ \\%%%%%%%%%%%%%%%%%%%%%%%%%%%%%%%%%%%%%%%%%%%%%%%%%%%%%%%%%%%%%%%
 	& $10^{-5}$ 	& 14&07 	& $<10^{-100}$ \\%%%%%%%%%%%%%%%%%%%%%%%%%%%%%%%%%%%%%%%%%%%%%%%%%%%%%%%%%%%%%%
 	& $10^{-6}$ 	& 137&82 	& $<10^{-100}$ \\%\hline%%%%%%%%%%%%%%%%%%%%%%%%%%%%%%%%%%%%%%%%%%%%%%%%%%%%%%%%%%%%%
	& $10^{-7}$ 	& 1378&33 & $<10^{-100}$ \\%\hline%%%%%%%%%%%%%%%%%%%%%%%%%%%%%%%%%%%%%%%%%%%%%%%%%%%%%%%%%%%%%

Exact &	--	& 0&19 	& 0.16 	\\
Exact' & -- 	& 0&17 	&  0.21%%%%%%%%%%%%%%%%%%%%%%%%%%%%%%%%%%%%%%%%%%%%%%%%%%%%%%%%%%%
%%%%%%%%%%%%%%%%%%%%%%%%%%%%%%%%%%%%%%%%%%%%%%%%%%%%%%%%%%%%%%%
\end{tabular} &
\begin{tabular}{c c r@{.}l c }
\multicolumn{5}{c}{$x = 0.5$}\\\hline
Method & $\gd$ & \multicolumn{2}{c}{Time (s)} & $p$-value\\\hline
\multirow{5}{*}{BISS}	%& $10^{-2}$ 	& 0&04 	& $5.9\times 10^{-70}$ 	\\%%%%%%%%%%%%%%%%%%%%%%%%%%%%%%%%%%%%%%%%%%%%%%%%%%%%%%%%%%
 	& $10^{-3}$ 	& 0&16 	& $1.2 \times 10^{-28}$ 	\\%%%%%%%%%%%%%%%%%%%%%%%%%%%%%%%%%%%%%%%%%%%%%%%%%%%%%%%%%%
	& $10^{-4}$ 	& 1&43 	& $9.0 \times 10^{-18}$ 	\\%%%%%%%%%%%%%%%%%%%%%%%%%%%%%%%%%%%%%%%%%%%%%%%%%%%%%%%%%%%%
	&  $10^{-5}$ 	& 14&08 	& $9.4\times 10^{-17}$ 	\\%%%%%%%%%%%%%%%%%%%%%%%%%%%%%%%%%%%%%%%%%%%%%%%%%%%%%%%%%
	& $10^{-6}$ 	& 138&23 	& $6.3\times 10^{-13}$ 	\\%%%%%%%%%%%%%%%%%%%%%%%%%%%%%%%%%%%%%%%%%%%%%%%%%%%%%%%%
	& $10^{-7}$ 	& 1368&33 	& $8.1\times 10^{-13}$ 	\\%%%%%%%%%%%%%%%%%%%%%%%%%%%%%%%%%%%%%%%%%%%%%%%%%%%%%%%%
	%\hline
Exact & -- 		& 0&19 	& 0.81 	\\
Exact' & -- 	& 0&17 	&  0.60	%%%%%%%%%%%%%%%%%%%%%%%%%%%%%%%%%%%%%%%%%%%%%%%%%%%%%%%%%%%
\end{tabular} \\\hline%%%%%%%%%%%%%%%%%%%%%%%%%%%%%%%%%%%%%%%%%%%%%%%%%%%%%%%%%%%%%%%%
 \\

\multicolumn{2}{c}{$t=5$}\\\hline
\begin{tabular}{c c r@{.}l c }
\multicolumn{5}{c}{$x = 0.01$}\\\hline
Method & $\gd$ & \multicolumn{2}{c}{Time (s)} & $p$-value\\\hline
\multirow{5}{*}{BISS} & $10^{-2}$ 	& 0&17 	& $<10^{-100}$ \\%%%%%%%%%%%%%%%%%%%%%%%%%%%%%%%%%%%%%%%%%%%%%%%%%%%%%%%%%%%%%%%%%
& $10^{-3}$ 	& 1&43 	& $<10^{-100}$ 	\\%%%%%%%%%%%%%%%%%%%%%%%%%%%%%%%%%%%%%%%%%%%%%%%%%%%%%%%%%%
& $10^{-4}$ 	& 14&01 	& $<10^{-100}$ 	\\%%%%%%%%%%%%%%%%%%%%%%%%%%%%%%%%%%%%%%%%%%%%%%%%%%%%%%%%%
	& $10^{-5}$ 	& 137&95 	& $<10^{-100}$ 	\\%%%%%%%%%%%%%%%%%%%%%%%%%%%%%%%%%%%%%%%%%%%%%%%%%%%%%%%%
& $10^{-6}$ 	& 1375&75 	& $<10^{-100}$ 	\\%\hline%%%%%%%%%%%%%%%%%%%%%%%%%%%%%%%%%%%%%%%%%%%%%%%%%%%%%%%
Exact & -- 		& 0&18 	& 0.58 	\\
Exact' & -- 	& 0&17 	& $1.2 \times 10^{-47}$%%%%%%%%%%%%%%%%%%%%%%%%%%%%%%%%%%%%%%%%%%%%%%%%%%%%%%%%%%%
%%%%%%%%%%%%%%%%%%%%%%%%%%%%%%%%%%%%%%%%%%%%%%%%%%%%%%%%%%%%%%%%%
\end{tabular} &
\begin{tabular}{c c r@{.}l c }
\multicolumn{5}{c}{$x = 0.5$}\\\hline
Method & $\gd$ & \multicolumn{2}{c}{Time (s)} & $p$-value\\\hline
\multirow{5}{*}{BISS} & $10^{-2}$ 	& 0&16 	& $<10^{-100}$ 	\\%%%%%%%%%%%%%%%%%%%%%%%%%%%%%%%%%%%%%%%%%%%%%%%%%%%%%%%%%%%
& $10^{-3}$ 	& 1&43 	& $<10^{-100}$ 	\\%%%%%%%%%%%%%%%%%%%%%%%%%%%%%%%%%%%%%%%%%%%%%%%%%%%%%%%%%%%
& $10^{-4}$ 	& 14&04 	& $<10^{-100}$ 	\\%%%%%%%%%%%%%%%%%%%%%%%%%%%%%%%%%%%%%%%%%%%%%%%%%%%%%%%%%%
& $10^{-5}$ 	& 138&09 	& $4.7\times 10^{-100}$ 	\\%%%%%%%%%%%%%%%%%%%%%%%%%%%%%%%%%%%%%%%%%%%%%%%%%%%%%%%%%
& $10^{-6}$ 	& 1378&54 	& $1.6\times 10^{-100}$ 	\\%\hline%%%%%%%%%%%%%%%%%%%%%%%%%%%%%%%%%%%%%%%%%%%%%%%%%%%%%%%%
Exact & --		& 0&18 	& 0.88 	\\
Exact' & -- 	& 0&17 	& 0.49 %%%%%%%%%%%%%%%%%%%%%%%%%%%%%%%%%%%%%%%%%%%%%%%%%%%%%%%%%%%
\end{tabular} \\\hline%%%%%%%%%%%%%%%%%%%%%%%%%%%%%%%%%%%%%%%%%%%%%%%%%%%%%%%%%%%%%%%%%%  
\end{tabular}}
\end{center}
\end{table}

\begin{figure}[t]
\begin{center}
\begin{tabular}{ccc}
 \includegraphics[width=0.4\textwidth]{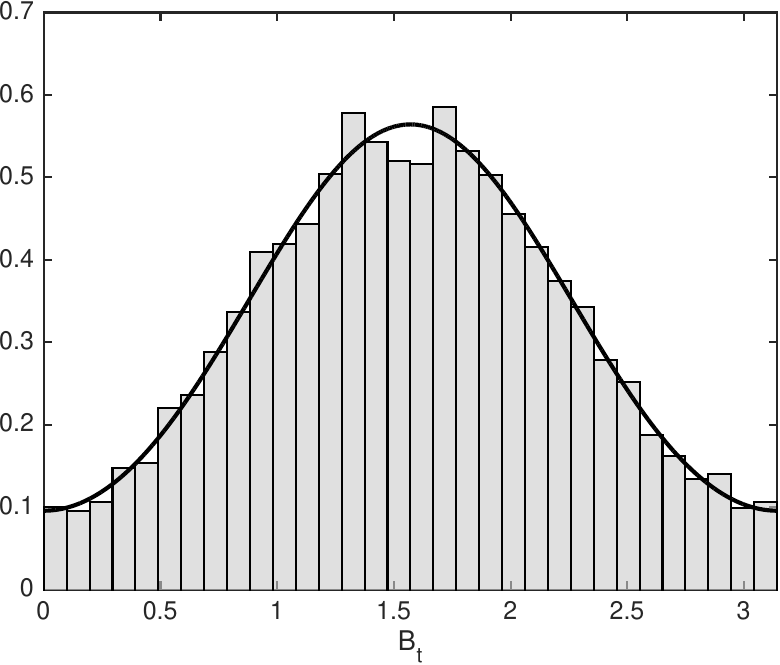} && \includegraphics[width=0.4\textwidth]{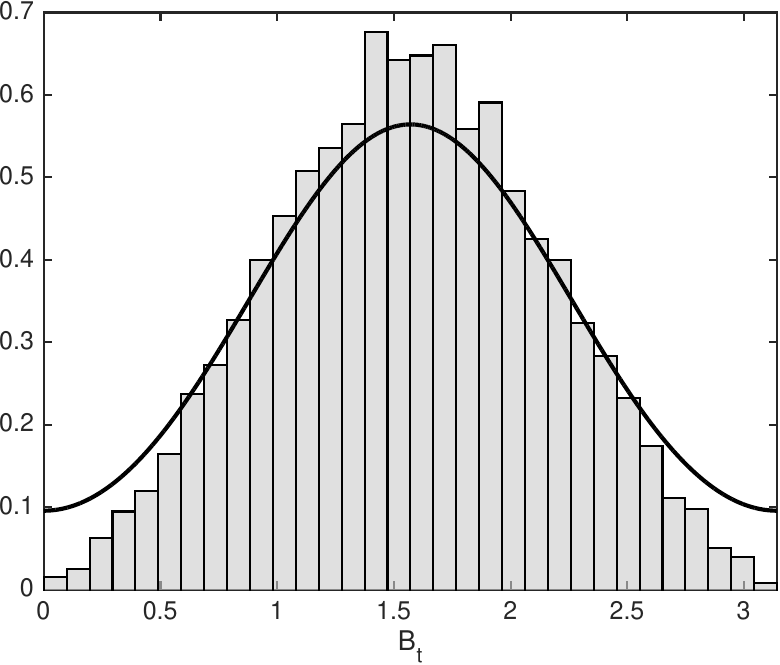}\\
 (a) && (b)
 \end{tabular}
 \end{center}
\caption{\label{fig:Euler-example}Histogram of 10,000 simulated points using (a) Algorithm \ref{alg:f} and (b) the Euler-type BISS algorithm of \citet{dan:etal:2012}. Parameters are $\gq_1 = \gq_2 = 1/2$, $x = 0.5$, $t = 0.5$, $\gd = 10^{-6}$. For visual clarity, plotted are samples of the driving reflecting Brownian motion $B_t = \arccos(1-2X_t)$ rather than $X_t$ (since the density function of $B_t$---shown as a solid line---is bounded and can be calculated; see main text).}
\end{figure}

As is evident from \tref{tab:neutralWFresults}, exact simulation strongly outperforms the BISS algorithm over almost all the parameter combinations investigated. Over a timescale of $t \gtrapprox 0.1$, errors in the Euler-type method accumulate sufficiently fast that the resulting samples are easy to reject in a K-S test. Even reducing the stepsize so that its running time is several orders of magnitude greater than the exact method provides only a modest improvement to the quality of the sample. Note also that the performance of the BISS algorithm deteriorates for paths started close to the boundary (compare $x=0.01$ with $x = 0.5$), whereas the exact method is indifferent to starting position. One reason $p$-values are persistently small for the BISS algorithm is that sample paths are constrained by construction to stay inside $[\gd/4, 1- \gd/4]$, yet the narrow region close to the boundaries is precisely where we expect to find much of the probability mass for many choices of parameters. An example of how this affects the resulting transition density is given in \fref{fig:Euler-example}. 
By contrast, in no application of the K-S test to samples from our method would we have rejected at level $0.05$ the hypothesis that they were drawn from the true distribution. Only over short timescales and away from the boundaries (\tref{tab:neutralWFresults}, $t \leq 0.05$ and $x=0.5$) do the two methods seem comparable. At $t=0.05$, the same computational investment as in the exact method but applied to the BISS method buys a stepsize of about $\gd = 10^{-4}$, which is adequate in the interior of the state space ($x=0.5$) but not near the boundaries ($x = 0.01$).

When $t\geq 0.05$ it is also possible to compare the exact method both with and without the approximation of \citet{gri:1984} (see p\pageref{page:1'}); the two versions exhibit similar running times and generate \emph{bona fide} samples from the true distribution (according to a K-S test) for moderate $t$. However, the approximate version deteriorates (as reported by the K-S $p$-value) for large $t$ (see \tref{tab:neutralWFresults}, $t = 5$), away from its asymptotic regime. Thus a suitable rule-of-thumb is to use the exact algorithm for $t \geq 0.05$ and its approximate version for $t < 0.05$, with the two methods in good agreement in their region of overlap in $t$.

We investigated several other choices of $x$ and $t$ with predictable results (not shown); for example, performance of the exact method seems unrelated to starting position $x$, while the BISS method deteriorates even further as $x \to 0$.

\section{Simulating the nonneutral Wright-Fisher process}
\label{sec:nonneutralWF}
In this section we develop an exact rejection algorithm for simulating from the Wright-Fisher diffusion \eqref{eq:WFSDE} with general drift. We make use of retrospective sampling techniques for the exact simulation of diffusions, which we first summarize.
\subsection{Overview of the exact algorithm}
\label{sec:EA}
Here we give a brief overview of the exact algorithm (EA) of \citet{bes:rob:2005}, \citet{bes:etal:2006:B, bes:etal:2008}, and \citet{pol:etal:2016}, and we refer the reader to those papers for further details. The EA returns a recipe for simulating the sample paths of a diffusion $X = (X_t)_{t\in[0,T]}$ satisfying the \sde{}
\begin{equation}
\label{eq:EA-SDE}
dX_t = \mu(X_t)dt + dB_t, \qquad X_0 = x_0, t \in [0,T],
\end{equation}
with $\mu$ assumed to satisfy the requirements for \eqref{eq:EA-SDE} to admit a unique, weak solution. Denote the law of such a process, our target, by $\bbQ_{x_0}$. The idea of the EA is to use Brownian motion started at $x_0$, whose law will be denoted $\bbW_{x_0}$, as the candidate process in a rejection sampling algorithm. The goal is then to write down the rejection probability, which is possible under the following assumptions:
\begin{itemize}
	\item[(A1)] The Radon-Nikod\'ym derivative of $\bbQ_{x_0}$ with respect to $\bbW_{x_0}$ exists and is given by Girsanov's formula,
	\begin{equation}
	\label{eq:Girsanov}
	\frac{d\bbQ_{x_0}}{d\bbW_{x_0}}(X) = \exp\left\{\int_0^T \mu(X_t) dX_t - \frac{1}{2}\int_0^T \mu^2(X_t)dt\right\},
	\end{equation}
	\item[(A2)] $\mu \in C^1$,
	\item[(A3)] $\phi(x) := \frac{1}{2}[\mu^2(x) + \mu'(x)]$ is bounded below, by $\phi^-$ say.
	\item[(A4)] $A(x) := \int_0^x \mu(z) dz$ is bounded above, by $A^+$ say.
\end{itemize}
Using (A1--A4) and It\^o's lemma, \eqref{eq:Girsanov} can be re-expressed as
\begin{align}
\label{eq:Girsanov2}
\frac{d\bbQ_{x_0}}{d\bbW_{x_0}}(X) \propto \exp\left\{A(X_T)-A^+\right\}\exp\left\{ - \int_0^T [\phi(X_t) - \phi^-] dt\right\}.
\end{align}
Written in this form, the right hand side of \eqref{eq:Girsanov2} is less than or equal to $1$, and therefore provides the required rejection probability. To make the accept/reject decision, it is necessary to construct an event occurring with probability \eqref{eq:Girsanov2}. This is easy to achieve given a realized sample path $(X_t)_{t\in[0,T]} \sim \bbW_{x_0}$, but obtaining such a path would require an infinite amount of computation. Instead, note that the right hand term in \eqref{eq:Girsanov2} is the probability that all points in a Poisson point process $\mathbf{\Phi} = \{ (t_j,  \psi_j) : j = 0,1,\ldots\}$ of unit rate on $[0,T] \times [0, \infty)$ lie in the epigraph of $t \mapsto [\phi(X_t) - \phi^-]$, and this event can be determined by simulating $X$ only at times $t_1, t_2, \ldots$. Thus, the following algorithm returns a (random) collection of skeleton points from $X \sim \bbQ_{x_0}$:
\\

\begin{algorithm}[H]
\DontPrintSemicolon
\Repeat{false}{
Simulate $\mathbf{\Phi}$, a Poisson point process on $[0,T] \times [0,\infty)$.\;
Simulate $U \sim \text{Uniform}[0,1]$.\;
Given $\mathbf{\Phi} = \{ (t_j,  \psi_j) : j = 0,1,\ldots\}$, simulate $B \sim \bbW_x$ at times $(t_j)_{j=0,1,\ldots}$ and at time $T$.\;
\If{$\phi(B_{t_j})-\phi^- \leq \psi_j$ for all $j = 0,1,\ldots$ and $U \leq \exp\{A(B_T) - A^+\}$}{\Return $\{(t_j, B_{t_j}) : j=0,1\,\ldots\} \cup \{(T,B_T)\}$.}
}
\caption{Exact algorithm for simulating the path of a diffusion process with law $\bbQ_x$.}\label{alg:EA}
\end{algorithm}
~\\

Once a skeleton has been accepted, further points can be filled in as required by sampling from Brownian bridges; no further reference to $\bbQ_{x}$ is necessary. If $\phi$ is bounded above, by $\phi^+$ say, then Algorithm \ref{alg:EA} can be implemented with finite computation by thinning $\mathbf{\Phi}$ to a Poisson Point process on $[0,T]\times [0,\phi^+-\phi^-]$; $|\mathbf{\Phi}|$ is then almost surely finite. [However, this requirement on $\phi$ can be relaxed \citep{bes:etal:2006:B, bes:etal:2008}.]

We remark that assumption (A4) is restrictive, and can in fact be removed by using a certain \emph{biased} Brownian motion as an alternative candidate; this also improves the efficiency of the algorithm \citep{bes:rob:2005}. We present the EA in the form above since an analogue of (A4) \emph{does} hold for the Wright-Fisher diffusion, and in any case a `biased Wright-Fisher diffusion' is not available.

\subsection{Exact algorithm for the Wright-Fisher diffusion}
As noted earlier, the requirements (A1--A4) need not hold when our target is the Wright-Fisher diffusion. However, related techniques can be used if the candidate process is chosen to be another Wright-Fisher diffusion, and in particular one with the same mutation parameters. Denote the target law $\WF_{\ggg,x_0}$ [with drift $\ggg$], and the candidate law $\WF_{\ga,x_0}$ [with drift \eqref{eq:neutral-drift}]. 
We will write $\ggg \in \cW\cF$ if there exists a drift $\ga$ of the form \eqref{eq:neutral-drift} such that the following hold:
\begin{itemize}
	\item[(WF1)] The Radon-Nikod\'ym derivative of $\WF_{\ggg,x_0}$ with respect to $\WF_{\ga,x_0}$ exists and is given by Girsanov's formula,
	\begin{multline}
	\label{eq:GirsanovWF}
	\frac{d\WF_{\ggg, x_0}}{d\WF_{\ga, x_0}}(X) = \exp\left\{\int_0^T \frac{\ggg(X_t) - \ga(X_t)}{X_t(1-X_t)} dX_t \right.\\
	\left. {}- \frac{1}{2}\int_0^T \frac{\ggg^2(X_t) - \ga^2(X_t)}{X_t(1-X_t)} dt\right\},
	\end{multline}
	\item[(WF2)] $\ggg$ is continuously differentiable on $(0,1)$.
	\item[(WF3)] $\widetilde{\phi}(x)$ is bounded on $(0,1)$: $\widetilde{\phi}^- \leq \widetilde{\phi}(x) \leq \widetilde{\phi}^+$, where
	\[
	\widetilde{\phi}(x) := \frac{1}{2}\left[\frac{\ggg^2(x) - \ga^2(x)}{x(1-x)} + \ggg'(x) - \ga'(x) - [\ggg(x) - \ga(x)]\frac{1-2x}{x(1-x)}\right].
	\]
	\item[(WF4)] $\widetilde{A}(x) := \int^x_{0} \frac{\ggg(z) - \ga(z)}{z(1-z)} dz$ is bounded above, by $\widetilde{A}^+$ say.
\end{itemize}
Specific conditions on $\alpha$ and $\ggg$ to satisfy (WF1) are detailed e.g.\ in Theorem 8.6.8 in \citet{oks:2003}. \citep[This theorem imposes some unduly restrictive conditions to ensure that the \sde{} has a unique weak solution, but this can be established for \eqref{eq:WFSDE} by other means;][Ch.\ 4.]{par:2009} 
 Following \sref{sec:EA}, we apply It\^o's lemma to $\widetilde{A}(x)$ to re-express \eqref{eq:GirsanovWF} as
\begin{align}
\label{eq:GirsanovWF2}
\frac{d\WF_{\ggg, x_0}}{d\WF_{\ga, x_0}}(X) \propto \exp\left\{\widetilde{A}(X_T)-\widetilde{A}^+\right\}\exp\left\{ - \int_0^T [\widetilde{\phi}(X_t) - \widetilde{\phi}^-] dt\right\}.
\end{align}
We recognize the rightmost term in \eqref{eq:GirsanovWF2} as the probability that no points in a Poisson point process on $[0,T]\times [0,\widetilde{\phi}^+-\widetilde{\phi}^-]$ lie in the epigraph of $t \mapsto \widetilde{\phi}(X_t) - \widetilde{\phi}^-$. Hence, Algorithm \ref{alg:WF-EA} returns exact samples from $\WF_{\ggg,x_0}$. Step \ref{alg:WF3} of Algorithm \ref{alg:WF-EA} is achieved via Algorithm \ref{alg:f}. Once a skeleton is accepted, further points can be filled in via Algorithm \ref{alg:fbridge}.

\begin{algorithm}[t]
\DontPrintSemicolon
\Repeat{false}{
Simulate $\mathbf{\Phi}$, a Poisson point process on $[0,T] \times [0,\widetilde{\phi}^+-\widetilde{\phi}^-]$.\;
Simulate $U \sim \text{Uniform}[0,1]$.\;
Given $\mathbf{\Phi} = \{ (t_j,  \psi_j) : j = 0,1,\ldots, J\}$, simulate $X \sim \WF_{\ga,x}$ at times $(t_j)_{j=0,1,\ldots,J}$ and at time $T$.\label{alg:WF3}\;
\If{$\widetilde{\phi}(X_{t_j})-\widetilde{\phi}^- \leq \psi_j$ for all $j = 0,1,\ldots, J$ and $U \leq \exp\{\widetilde{A}(X_T) - \widetilde{A}^+\}$\nllabel{eAtest}}{\Return $\{(t_j, X_{t_j}) : j=0,1\,\ldots,J\} \cup \{(T,X_T)\}$.}
}
\caption{Exact algorithm for simulating the path of a diffusion process with law $\WF_{\ggg,x}$.}\label{alg:WF-EA}
\end{algorithm}

\subsection{A class of drifts for which exact simulation is possible}
The class $\cW\cF$ defines the set of drifts for which exact simulation is possible. 
Here we show that $\cW\cF$ contains all the most popular population genetics diffusion processes with mutation and selection (including frequency-dependent selection), whose drift admits the general form
\begin{equation}
\gamma(x)=\alpha(x)+ x(1-x) \eta(x),
\label{eq:coopbob-drift}
\end{equation}
where $\alpha$ is as in \eqref{eq:neutral-drift} for some $\theta_1,\theta_2>0$ and $\eta$ is a reasonably regular function of $x$. The case of diploid selection \eqref{eq:selection-drift} is recovered by setting $\eta(x)\propto (x+h(1-2x))$. For general $\eta$ the properties of diffusion processes with drift \eqref{eq:coopbob-drift} and of the corresponding genealogies are studied, among others, by \cite{coo:gri:2004}. 

\begin{proposition}
\label{prop:coopbob}
Any drift of the form \eqref{eq:coopbob-drift}, for $\alpha$ as in \eqref{eq:neutral-drift} for some $\theta_1,\theta_2>0$ and for $\eta$ continuously differentiable in $[0,1]$, is a member of $\cW\cF$.
\end{proposition}
It is worth noting that the additive structure in the drift $\gamma$ (that is, with a selection component added to the mutation component $\alpha$) is a widely accepted and theoretically justified property of all population genetics diffusion models \citep[see e.g.\ the discussion in][p186--187]{kar:tay:1981}.

\subsubsection{Example: Wright-Fisher diffusion with diploid selection}
By \propref{prop:coopbob}, the drift $\gb$ [eq.~\eqref{eq:selection-drift}] satisfies $\gb \in \cW\cF$. In fact, in the haploid case ($h = 1/2$), there is much simplification: $\widetilde{\phi}$ is a quadratic polynomial for which analytic bounds are available, and $\widetilde{A}(x) = \gs x/2$. 
We implemented our exact algorithm on this model, and investigated its performance by considering several combinations of parameters; results are shown in \tref{tab:nonneutralWFresults}. For moderate selection ($|\sigma| = 1$) the algorithm is extremely efficient, with only slightly more than one candidate needed per acceptance. Furthermore, most simulations resulted in zero Poisson points. These results are quite robust to the length of the path $t$, the initial frequency $x$, and the sign of the selection parameter. For stronger selection ($\sigma = 10$) we observe some deterioration in efficiency because of the greater mismatch between candidate and target paths---to the extent that simulation of paths of length $t = 5.0$ became prohibitive. Nonetheless, it is still feasible to simulate a collection of shorter paths in a few seconds (and to string these together to construct longer paths, if necessary).

\begin{table}[!t]
\caption{\label{tab:nonneutralWFresults}Performance of Algorithm \ref{alg:WF-EA} applied to the Wright-Fisher diffusion with symmetric mutation and additive selection. Each row reports means (per accepted path) across a simulation to generate 1,000 accepted paths. Paths were initiated at $X_0 = x$ and run for time $t$, with mutation parameters $\theta_1=\theta_2 = 0.01$ and selection parameter $\sigma$ (and $h = 0.5$). Reported are the total numbers (per \emph{accepted} path) of: attempts, Poisson points simulated, coefficients $b^{(t,\theta)}_k(m)$ needed, random variables generated (that is, the aggregate of all draws from underlying constituent distributions: uniforms, betas, and so on), the number of times the simulation resorted to the approximation of \thmref{thm:gri:1984}, and the running time in milliseconds.} 
\begin{center}
\begin{tabular}{r@{.}l r@{.}l r@{.}l r@{.}l r@{.}l r@{.}l r@{.}l r@{.}l r@{.}l r@{.}l r}
%\multicolumn{15}{c}{$\sigma = -1$}\\
\multicolumn{15}{c}{$\sigma = 1$}\\
\hline
\multicolumn{2}{c}{} &  \multicolumn{2}{c}{} & \multicolumn{2}{c}{} & \multicolumn{2}{r}{Poisson} & \multicolumn{2}{c}{} & \multicolumn{2}{r}{Random} & \multicolumn{2}{c}{} & \multicolumn{2}{c}{} \\%& \multicolumn{2}{c}{}\\
\multicolumn{2}{c}{$t$} &  \multicolumn{2}{c}{$x$} & \multicolumn{2}{c}{Attempts} & \multicolumn{2}{r}{points} & \multicolumn{2}{r}{Coefficients} & \multicolumn{2}{r}{variables} & \multicolumn{2}{r}{G1984} & \multicolumn{2}{c}{Time (ms)}\\
\hline
%	0&1	&	0&5	&	1&3	&	0&0	&	118&4	&	7&5	&	0&0	&	2&0	\\
%	0&1	&	0&25	&	1&1	&	0&0	&	105&6	&	6&6	&	0&0	&	2&0	\\
%	0&1	&	0&01	&	1&0	&	0&0	&	91&1	&	6&0	&	0&0	&	2&0	\\
%	0&5	&	0&5	&	1&2	&	0&0	&	18&6	&	7&3	&	0&0	&	1&0	\\
%	0&5	&	0&25	&	1&1	&	0&0	&	17&1	&	6&8	&	0&0	&	1&0	\\
%	0&5	&	0&01	&	1&0	&	0&0	&	15&0	&	6&1	&	0&0	&	0&0	\\
%%	1&0	&	0&5	&	1&3	&	0&0	&	10&9	&	7&4	&	0&0	&	0&0	\\
%%	1&0	&	0&25	&	1&1	&	0&0	&	9&6	&	6&8	&	0&0	&	0&0	\\
%%	1&0	&	0&01	&	1&0	&	0&0	&	9&4	&	6&2	&	0&0	&	1&0	\\
%	5&0	&	0&5	&	1&3	&	0&2	&	5&6	&	8&8	&	0&0	&	0&0	\\
%	5&0	&	0&25	&	1&2	&	0&2	&	6&3	&	7&8	&	0&0	&	0&0	\\
%	5&0	&	0&01	&	1&0	&	0&2	&	4&9	&	7&2	&	0&0	&	0&0	\\
%	\hline \multicolumn{15}{c}{}\\
%	\multicolumn{15}{c}{$\sigma = 1$}\\ \hline
	0&1	&	0&5	&	1&27	&	0&00	&	116&07	&	7&37	&	0&00	&	0&007	\\
	0&1	&	0&25	&	1&40	&	0&00	&	132&87	&	8&01	&	0&00	&	0&008\\
	0&1	&	0&01	&	1&58	&	0&01	&	141&34	&	8&91	&	0&01	&	0&009\\
	0&5	&	0&5	&	1&23	&	0&03	&	18&65 &	7&22	&	0&00	&	0&003\\
	0&5	&	0&25	&	1&48	&	0&03	&	21&60 &	8&45	&	0&01	&	0&003	\\
	0&5	&	0&01	&	1&58	&	0&03	&	23&84 &	9&02	&	0&01	&	0&003	\\
	5&0	&	0&5	&	1&29	&	0&24	&	5&23	&	8&58	&	0&00	&	0&002	\\
	5&0	&	0&25	&	1&49	&	0&27	&	6&18	&	9&64	&	0&01	&	0&002	\\
	5&0	&	0&01	&	1&67	&	0&28	&	7&36	&	10&79&	0&01	&	0&002	\\
	\hline \multicolumn{15}{c}{}\\
	\multicolumn{15}{c}{$\sigma = 10$}\\ \hline
	0&1	&	0&5	&	11&83	&	3&72		&	995&92	&	62&68	&	1&83	&	0&062	\\
	0&1	&	0&25	&	41&69	&	12&97	&	3714&82	&	224&95	&	7&86	&	0&225	\\
	0&1	&	0&01	&	145&73	&	45&96	&	14937&52	&	856&21	&	45&88	&	0&879	\\
	0&5	&	0&5	&	13&16	&	20&96	&	641&52	&	109&00	&	2&47	&	0&054\\
	0&5	&	0&25	&	43&82	&	69&05	&	2729&80	&	399&95	&	10&89	&	0&214	\\
	0&5	&	0&01	&	149&21	&	235&34	&	17044&43	&	1869&54	&	71&54	&	1&185	\\
  \hline
\end{tabular}
\end{center}
\end{table}

To make this observation more precise, \citet[Proposition 3]{bes:etal:2006:B} obtained an explicit upper bound on the expected number of Poisson points required of Algorithm \ref{alg:EA}, and hence on the computational complexity of the algorithm. A careful reading of their result shows that it relies on the existence of the bounds $\phi^\pm$ but does not depend on the laws of the diffusions involved, and carries over easily to the Wright-Fisher case. We therefore omit a proof of the following.
\begin{proposition}
\label{prop:nonneutralcomplexity}
Let $N^{(T,\gq)}$ denote the number of Poisson points required until the first accepted path and suppose $\widetilde{A}^-$ is a lower bound on $\widetilde{A}(x)$. Then
\[
\bbE[N^{(t,\gq)}] \leq (\widetilde{\phi}^+ - \widetilde{\phi}^-)Te^{(\widetilde{\phi}^+ - \widetilde{\phi}^-)T + \widetilde{A}^+-\widetilde{A}^-}.
\]
\end{proposition}
An immediate consequence of \propref{prop:nonneutralcomplexity} is that the complexity of simulating a path of length $KT$ is $O((\widetilde{\phi}^+ - \widetilde{\phi}^-)KTe^{(\widetilde{\phi}^+ - \widetilde{\phi}^-)KT + \widetilde{A}^+-\widetilde{A}^-})$ as $KT \to \infty$. However, superior performance can be achieved by splitting the problem into $K$ separate simulations; the complexity is then $O((\widetilde{\phi}^+ - \widetilde{\phi}^-)KTe^{(\widetilde{\phi}^+ - \widetilde{\phi}^-)T + \widetilde{A}^+-\widetilde{A}^-})$, which is linear in $K$ as in \citet{bes:etal:2006:B}.

As this is a statement about the complexity of the algorithm as the path length increases, it continues to hold even if we account for the increasing cost associated with each Poisson point as $T \to 0$, as quantified by \propref{prop:neutralcomplexity}. In practice one might wish to optimize the choice of $K$ and $T$ for a given path length $KT$. In our application, we must be prepared to introduce an additional constraint in order to fix $T$ some distance away from $0$ (and, as we recommend above, the choice $T \geq 0.05$ seems adequate).

\section{Simulating a nonneutral Wright-Fisher bridge}
\label{sec:nonneutralbridge}
For completeness, here we provide an algorithm for simulating a nonneutral Wright-Fisher \emph{bridge} (Algorithm \ref{alg:WF-EA-bridge}). This follows immediately from the previous sections; the only modification is to note that the appropriate candidate process is the corresponding neutral bridge, which can be simulated via Algorithm \ref{alg:fbridge}. The rest follows exactly as in the Brownian case; see \citet[Section 6.2]{bes:etal:2006:B} for details.

\begin{algorithm}[t]
\DontPrintSemicolon
\Repeat{false}{
Simulate $\mathbf{\Phi}$, a Poisson point process on $[0,T] \times [0,\widetilde{\phi}^+-\widetilde{\phi}^-]$.\;
Given $\mathbf{\Phi} = \{ (t_j,  \psi_j) : j = 0,1,\ldots, J\}$, simulate $X \sim (\WF_{\ga,x}\mid X_T = y)$ at times $(t_j)_{j=0,1,\ldots,J}$.\;
\If{$\widetilde{\phi}(X_{t_j})-\widetilde{\phi}^- \leq \psi_j$ for all $j = 0,1,\ldots, J$}{\Return $\{(t_j, X_{t_j}) : j=0,1\,\ldots,J\} \cup \{(T,y)\}$.}
}
\caption{Exact algorithm for simulating the path of a diffusion process with law $\WF_{\ggg,x}$ conditioned on $X_T = y$.}\label{alg:WF-EA-bridge}
\end{algorithm}

\section{Discussion}
\label{sec:discussion}
In this paper we have shown how to simulate exactly from the scalar Wright-Fisher diffusion, as well as a number of important and closely related processes: these include the ancestral process of an $\infty$-leaf Kingman coalescent tree, the Fleming-Viot process with parent-independent mutation, the nonneutral Wright-Fisher diffusion, and neutral and nonneutral Wright-Fisher bridges. Some interesting open problems remain, including mutation operators which do not lead to reversible diffusions, and the problem of sampling from $(d-1)$-dimensional ($2 < d\leq\infty$) Wright-Fisher bridges.

It is also important to notice that, in order to employ the machinery proposed in this paper, the mutation parameters $\theta_1,\theta_2$ in the $\alpha$-component of a general drift $\gamma$ need to be both positive: models of the form  \eqref{eq:coopbob-drift} with no mutation ($\alpha=0$) or with one-directional mutation (with only one mutation parameter positive and the other null) have at least one absorbing boundary, and therefore there cannot be absolute continuity with respect to a stationary Wright-Fisher diffusion with a transition density expansion of the form of \eqref{eq:bridge-mixture}. For such cases, series expansions are in fact available with a structure similar to \eqref{eq:transition-dual} \citep[see e.g.][]{eth:gri:1993} and we believe it should be relatively simple to adapt our method to encompass selection models with absorbing boundaries, a goal we do not pursue here. 
For choices of drifts $\gamma$ more general than \eqref{eq:coopbob-drift}, arising possibly in models beyond population genetics, it is harder to specify conditions for (WF1) verifiable in a simple way by inspection of $\gamma.$ The determination of which drift functions $\gamma$ guarantee (assuming identical diffusion coefficients) absolute continuity with respect to a Wright-Fisher process, seems to be, to our knowledge, an open problem. We expect that most of what is affected by the drift pertains to the behaviour of the process at the boundaries. The problem, however, might be delicate and not just limited to making sure that $\gamma$ prevents the boundaries from being absorbing: for example, if for the diffusion with drift $\gamma$  the boundaries are entrance or reflecting, the rate of escape from the boundaries might still make its sample paths qualitatively different from those of any Wright-Fisher diffusion, even within the same boundary regime, respectively entrance or reflecting. To support this conjecture, we remark that the very same circumstances induce mutual singularity in squared Bessel processes (whose diffusion coefficient is $\sqrt x$ hence quite similar near zero to the Wright-Fisher diffusion): it is well-known that any two squared Bessel processes starting at 0 are mutually singular whenever their drifts differ, even if they are both within the same boundary regime \citep[see][Lemma (2.1) and references therein]{pit:yor:1981}. 

In a wider perspective, we believe that the approach proposed here might serve as a template for developing new techniques for sampling exactly from diffusion processes by means of non-Brownian bridges whose transition function admits a transparent transition function expansion.

%%%
\section{Proofs}
\label{sec:proofs}
{\bf Proof of \propref{prop:qm}.} First suppose $m > 0$.\\
(i) Note that
\begin{equation}
\label{eq:ratio}
\frac{b^{(t,\theta)}_{k+1}(m)}{b^{(t,\theta)}_{k}(m)} = \frac{\theta + m + k - 1}{k-m+1}\cdot\frac{\theta + 2k + 1}{\theta + 2k - 1}e^{-\frac{(2k+\theta)t}{2}} =: f^\theta_m(k)e^{-\frac{(2k+\theta)t}{2}},
\end{equation}
say. Treat $f^\theta_m(k)$ as having domain $\bbR$; it then suffices to show that $(f^{\theta}_m)'(k) < 0$ for all sufficiently large $k$. Then the right hand side of \eqref{eq:ratio} is subsequently decreasing in $k$ monotonically to $0$. Part (i) follows for the finite $k$ ($= m+i$) for which the right hand side of \eqref{eq:ratio} drops below $1$. Routine calculations show that $(f^\theta_m)'(k) < 0$ for all $k > (\sqrt{2(m-1) + \theta} - \theta)/2$.\\
(ii) Note that
\[
\frac{\sqrt{2(m-1) + \theta} - \theta}{2} < \sqrt{\frac{m-1}{2}} + \frac{\sqrt{\theta} - \theta}{2} <  \sqrt{\frac{m-1}{2}} + \frac{1}{8} < m,
\]
so in fact $(f^\theta_m)'(k) < 0$ for \emph{all} $k \geq m$, and thus as soon as $b^{(t,\theta)}_{k+1}(m) < b^{(t,\theta)}_{k}(m)$ for some $k$, it must also hold for all subsequent $k$.\\
(iii) The right hand side of \eqref{eq:ratio} can be made independent of $m$ by noting that
\begin{equation}
\label{eq:finequality}
f^\theta_m(k) < f^\theta_k(k) = \theta + 2k + 1.
\end{equation}
Thus for $\Ca^{(t,\gq)}_m = 0$ to hold we need $m$ to exceed the upper of the two solutions on $\bbR$ of
\[
(\theta + 2k +1)e^{-(2k+\theta)t/2} = 1.
\]
The definition of $\Cb^{(t,\gq)}_0$ is one way to express this condition, since the maximum of $(\theta + 2k +1)e^{-(2k+\theta)t/2}$ which separates the two solutions is at $k = \left(\frac{1}{t} - \frac{\theta + 1}{2}\right)$. 
Finally, consider the special case $m=0$. If $\theta > 1$ then similar arguments as in (i--ii) above continue to hold. However, if $\theta \leq 1$ then in fact $(f^\theta_0)'(k) > 0$ for all $k$, with $f_0^\theta(k)$ continuous on $k \geq 1$. But then $f^\theta_0(k) < f^\theta_0(\infty) = 1$ for $k \geq 1$, so $f_0^\theta(k)e^{-(2k+\theta)t/2} < 1$ for all $k \geq 1$ and hence (i--ii) still hold, with $\Ca^{(t,\gq)}_0 \leq 1$. \qed
\\

\noindent {\bf Proof of \propref{prop:WFbridge}.} This follows by substituting \eqref{eq:transition-dual} into \eqref{eq:bridge-transition}, multiplying by $B(\gq_1 + l +j, \gq_2 + m-l + k-j)/B(\gq_1 + l +j, \gq_2 + m-l + k-j)$, and rearranging.\qed
\\

\noindent {\bf Proof of \lemmaref{lem:beta}}\footnote{This proof corrects an error in earlier versions, and follows an argument presented in \citet{gar:etal:arxiv}.}. First suppose $l \leq \lfloor mz\rfloor$. Then, using $\Gamma(x+1) = x\Gamma(x)$,
\begin{align*}
\MoveEqLeft{\bbP(L_{m+1} = l)\cD_{\gq_1 + l,\gq_2 + m+1-l}(z)}\\
&= \left[\frac{m+1}{m+1-l}(1-x)\frac{\gq+m}{\gq_2 + m-l}(1-z)\right]\bbP(L_m = l)\cD_{\gq_1 + l, \gq_2 + m-l}(z),\\
&\leq \left[\frac{m+1}{m+1-mz}\frac{\gq+m}{\gq_2 + m-mz}\right](1-x)(1-z)\bbP(L_m = l)\cD_{\gq_1 + l, \gq_2 + m-l}(z).
\end{align*}
Now, to bound the term $g(m) := \frac{m+1}{m+1-mz}\cdot\frac{\theta + m}{\theta_2 + m - mz}$ in brackets we write $g(m) = a(m)b(m)$ and note that $a(m) := \frac{m+1}{1+m(1-z)}$ is increasing in $m$, so that $a(m) \leq a(\infty) = (1-z)^{-1}$. Further, for $m \geq 1$,
\[
b(m) := \frac{\theta + m}{\theta_2 + m - mz} \leq \frac{\theta + m}{m(1-z)} = \frac{1+\frac{\theta}{m}}{1-z} \leq \frac{1+\theta}{1-z}.
\]
Hence $g(m) \leq g(0) \vee a(\infty)\frac{1+\theta}{1-z} = \frac{\theta}{\theta_2}\vee \frac{1+\theta}{(1-z)^2}$, and so
\begin{multline*}
\bbP(L_{m+1} = l)\cD_{\gq_1 + l,\gq_2 + m+1-l}(z)\\ \leq \left[\frac{\theta}{\theta_2}(1-z)\vee \frac{1+\theta}{(1-z)}\right](1-x)\bbP(L_m = l)\cD_{\gq_1 + l, \gq_2 + m-l}(z).
\end{multline*}
Hence, summing over $l=0,\ldots, \lfloor mz \rfloor$,
\begin{multline}
\label{eq:partitionmz1}
\sum_{l=0}^{\lfloor mz \rfloor} \bbP(L_{m+1} = l)\cD_{\gq_1 + l,\gq_2 + m+1-l}(z) \\ \leq  \left[\frac{\theta}{\theta_2}(1-z)\vee \frac{1+\theta}{(1-z)}\right](1-x)\sum_{l=0}^{\lfloor mz\rfloor} \bbP(L_m = l)\cD_{\gq_1 + l, \gq_2 + m-l}(z).
\end{multline}
We employ a similar argument for $l \geq \lfloor mz \rfloor$:
\begin{align*}
\MoveEqLeft{\bbP(L_{m+1} = l+1)\cD_{\gq_1 + (l+1),\gq_2 + (m+1)-(l+1)}(z)}\\
&= \left[\frac{m+1}{l+1}\frac{\gq+m}{\gq_1 + l}\right]xz\bbP(L_m = l)\cD_{\gq_1 + l, \gq_2 + m-l}(z).
\end{align*}
To find an upper bound for
\[
h(l,m) := \frac{m+1}{l+1}\cdot \frac{\theta+m}{\theta_1+l},
\]
we consider the cases (i) $mz > 1$ and (ii) $mz \leq 1$ in turn. For (i) we have
\[
h(l,m) \leq h(mz-1,m) = \frac{m+1}{mz}\cdot\frac{\theta+m}{\theta_1+mz-1} \leq \left(1+\frac{1}{z}\right)\cdot\frac{\theta+m}{\theta_1+mz-1} =: c(m).
\]
It is straightforward to verify that the sign of $c'(m)$ is independent of $m$, and thus
\[
h(l,m) \leq c(z^{-1}) \vee c(\infty) = \left(1+\frac{1}{z}\right)\left[\frac{z\theta+1}{z\theta_1} \vee \frac{1}{z}\right].
\]
For case (ii), we have
\[
h(l,m) \leq h(0,m) = (m+1)\cdot\frac{\theta+m}{\theta_1} \leq \left(1+\frac{1}{z}\right)\cdot\frac{z\theta + 1}{z\theta_1}.
\]
Thus, in both cases we have
\[
h(l,m) \leq \left(1+\frac{1}{z}\right)\left[\frac{z\theta+1}{z\theta_1} \vee \frac{1}{z}\right],
\]
and hence
\begin{multline*}
\bbP(L_{m+1} = l+1)\cD_{\gq_1 + (l+1),\gq_2 + m+1-(l+1)}(z)\\ \leq  \left(1+\frac{1}{z}\right)\left[\frac{z\theta + 1}{\theta_1} \vee 1\right]x\bbP(L_m = l)\cD_{\gq_1 + l, \gq_2 + m-l}(z).
\end{multline*}
(This time it is crucial we compare $L_{m+1} = l+1$ with $L_m = l$, rather than with $L_m = l+1$.) Thus
\begin{multline}
\label{eq:partitionmz2}
\sum_{l=\lfloor mz \rfloor+1}^{m+1} \bbP(L_{m+1} = l)\cD_{\gq_1 + l,\gq_2 + m+1-l}(z) \\ \leq \left(1+\frac{1}{z}\right)\left[\frac{z\theta + 1}{\theta_1} \vee 1\right]x\sum_{l=\lfloor mz\rfloor}^m \bbP(L_m = l)\cD_{\gq_1 + l, \gq_2 + m-l}(z).
\end{multline}
Finally, sum \eqref{eq:partitionmz1} and \eqref{eq:partitionmz2} to yield \eqref{eq:beta}, noting that the overlap on the right-hand side at $l = \lfloor mz \rfloor$ necessitates the given definition of $\K^{(x,z)}$ (summing the two contributions from $l = \lfloor mz\rfloor$). \qed
\\

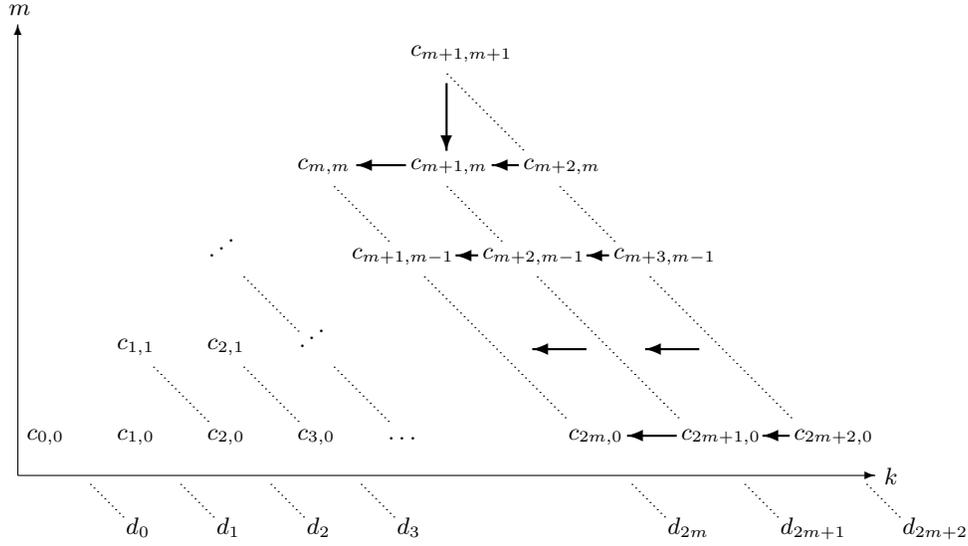
\begin{figure}[t]
\centering
\setlength{\unitlength}{1.2cm}
\begin{picture}(11,6.3)(0.9,0.25)
	\thinlines
        %% axis
	\put(0.9,0.9){\vector(0,1){5}}
	\put(0.9,0.9){\vector(1,0){9.5}}
	\put(0.8,6.0){$m$}
	\put(10.5,0.8){$k$}
	
%	\put(0.7,5.3){$4$}
%	\put(0.7,4.3){$3$}
%	\put(0.7,3.3){$2$}
%	\put(0.7,2.3){$1$}
%	\put(0.7,1.3){$0$}

\put(4,4.3){$c_{m,m}$}
\put(4.6,3.3){$c_{m+1,m-1}$}
%\put(7,3.3){$\ddots$}
%\put(8,2.3){$\ddots$}
\put(7,1.3){$c_{2m,0}$}

\put(5.25,4.3){$c_{m+1,m}$}
\put(6.05,3.3){$c_{m+2,m-1}$}
%\put(8,3.3){$\ddots$}
%\put(9,2.3){$\ddots$}
\put(8.25,1.3){$c_{2m+1,0}$}

\put(5.25,5.55){$c_{m+1,m+1}$}
\put(6.5,4.3){$c_{m+2,m}$}
\put(7.5,3.3){$c_{m+3,m-1}$}
%\put(8,4.3){$\ddots$}
%\put(9,3.3){$\ddots$}
%\put(10,2.3){$\ddots$}
\put(9.5,1.3){$c_{2m+2,0}$}

\put(3,3.3){$\iddots$}

\put(2,2.3){$c_{1,1}$}
\put(3,2.3){$c_{2,1}$}
\put(4,2.3){$\iddots$}

\put(1,1.3){$c_{0,0}$}
\put(2,1.3){$c_{1,0}$}
\put(3,1.3){$c_{2,0}$}
\put(4,1.3){$c_{3,0}$}
\put(5,1.3){$\ldots$}

%\put(1,0.3){$0$}
%\put(2,0.3){$1$}
%\put(3,0.3){$2$}
%\put(4,0.3){$3$}
%\put(5,0.3){$4$}
%\put(6,0.3){$\ldots$}

\multiput(2.4,2.1)(0.04,-0.04){16}{\line(1,0){0.02}}
\multiput(3.4,2.1)(0.04,-0.04){16}{\line(1,0){0.02}}
\multiput(4.4,2.1)(0.04,-0.04){16}{\line(1,0){0.02}}
\multiput(3.4,3.1)(0.04,-0.04){16}{\line(1,0){0.02}}

\multiput(4.4,4.1)(0.04,-0.04){16}{\line(1,0){0.02}}
\multiput(5.4,3.1)(0.04,-0.04){40}{\line(1,0){0.02}}

\multiput(5.65,4.1)(0.04,-0.04){16}{\line(1,0){0.02}}
\multiput(6.65,3.1)(0.04,-0.04){40}{\line(1,0){0.02}}

\multiput(5.65,5.35)(0.04,-0.04){22}{\line(1,0){0.02}}
\multiput(6.9,4.1)(0.04,-0.04){16}{\line(1,0){0.02}}
\multiput(7.9,3.1)(0.04,-0.04){40}{\line(1,0){0.02}}

\multiput(7.7,0.8)(0.04,-0.04){10}{\line(1,0){0.02}}
\multiput(8.95,0.8)(0.04,-0.04){10}{\line(1,0){0.02}}
\multiput(10.3,0.8)(0.04,-0.04){10}{\line(1,0){0.02}}

\multiput(1.7,0.8)(0.04,-0.04){10}{\line(1,0){0.02}}
\multiput(2.7,0.8)(0.04,-0.04){10}{\line(1,0){0.02}}
\multiput(3.7,0.8)(0.04,-0.04){10}{\line(1,0){0.02}}
\multiput(4.7,0.8)(0.04,-0.04){10}{\line(1,0){0.02}}
\put(2.1,0.25){$d_{0}$}
\put(3.1,0.25){$d_{1}$}
\put(4.1,0.25){$d_{2}$}
\put(5.1,0.25){$d_{3}$}

\put(8.1,0.25){$d_{2m}$}
\put(9.35,0.25){$d_{2m+1}$}
\put(10.7,0.25){$d_{2m+2}$}

\thicklines
\put(5.2,4.34){\vector(-1,0){0.55}}
\put(6.0,3.34){\vector(-1,0){0.25}}
\put(7.2,2.3){\vector(-1,0){0.6}}
\put(8.2,1.34){\vector(-1,0){0.55}}

\put(6.45,4.34){\vector(-1,0){0.3}}
\put(7.45,3.34){\vector(-1,0){0.25}}
\put(8.45,2.3){\vector(-1,0){0.6}}
\put(9.45,1.34){\vector(-1,0){0.3}}

\put(5.65,5.25){\vector(0,-1){0.75}}

\end{picture}
\caption{\label{fig:dm}Computation of $\cf{k}{m}$. The sum of each antidiagonals (dashed) defines the sequence $(d_i)_{i=0,1,\ldots}$. To show that $d_{i+1} < d_i$ terms are paired off as shown by the arrows; that is, the coefficient at the head of a set of arrows is greater in magnitude than the sum of the terms at its tails.}
\end{figure}

\noindent {\bf Proof of \propref{prop:lattice}.} First, since $m \geq \D^{(t,\gq)}$ we know $j \geq \Ca^{(t,\gq)}_{m-j}$ for $j=0,\ldots, m$ and hence by \propref{prop:qm} that $b_{m+j+1}^{(t,\gq)}(m-j) < b_{m+j}^{(t,\gq)}(m-j)$. Now multiply this inequality by $\bbE[\cD_{\gq_1 + L_{m-j}, \gq_2 + m-j - L_{m-j}}(z)]$ to yield
\begin{equation}
\label{eq:c}
\cf{m+j+1}{m-j} < \cf{m+j}{m-j}.
\end{equation}
Thus, summing over $j=0,1,\ldots, m$,
\begin{equation}
\label{eq:m-even}
\sum_{j=0}^m \cf{m+1+j}{m-j} < \sum_{j=0}^m \cf{m+j}{m-j},
\end{equation}
which says precisely that $d_{2m+1} < d_{2m}$. We also need to show that $d_{2m+2} < d_{2m+1}$, but this case is more subtle since the left hand side is a sum over one more term than the right. Instead, we will increment the first index in \eqref{eq:c} and then sum over $j=1,\ldots, m$:
\begin{equation}
\label{eq:c2}
\sum_{j=1}^{m} \cf{m+2+j}{m-j} < \sum_{j=1}^{m} \cf{m+1+j}{m-j}.
\end{equation}
It then suffices to show
\begin{equation}
\label{eq:suffices}
\cf{m+1}{m+1} + \cf{m+2}{m} < \cf{m+1}{m},
\end{equation}
for if we sum \eqref{eq:c2} and \eqref{eq:suffices} we obtain $d_{2m+2} < d_{2m+1}$ as required (\fref{fig:dm}).  To show \eqref{eq:suffices}, first note that
\[
\frac{\cf{k+1}{m}}{\cf{k}{m}} = \frac{b^{(t,\theta)}_{k+1}(m)}{b^{(t,\theta)}_{k}(m)} = f^\theta_m(k)e^{-\frac{(2k+\theta)t}{2}} \leq (\theta + 2k + 1)e^{-\frac{(2k+\theta)t}{2}},
\]
with $f^\theta_m(k)$ defined as in \eqref{eq:ratio} and the inequality following from \eqref{eq:finequality}. Hence, choosing $k = m+1$ and noting that $m \geq C_\ge^{(t,\gq)}$,
\begin{equation}
\label{eq:c-firsthalf}
\cf{m+2}{m} \leq (\theta + 2k + 1)e^{-\frac{(2k+\theta)t}{2}}\cf{m+1}{m} < (1-\ge)\cf{m+1}{m}.
\end{equation}
Second, note that
\begin{align}
\frac{\cf{m+1}{m+1}}{\cf{m+1}{m}} &= \frac{1}{m+1}\frac{\gq + 2m}{\gq + m}\frac{\bbE[\cD_{\gq_1 + L_{m+1},\gq_2 + m+1 - L_{m+1}}(z)]}{\bbE[\cD_{\gq_1 + L_{m},\gq_2 + m - L_{m}}(z)]}, \notag\\
&\leq \frac{1}{m+1}\cdot 2\frac{\bbE[\cD_{\gq_1 + L_{m+1},\gq_2 + m+1 - L_{m+1}}(z)]}{\bbE[\cD_{\gq_1 + L_{m},\gq_2 + m - L_{m}}(z)]}, \notag\\
&< \ge, \label{eq:c-secondhalf}
\end{align}
using \lemmaref{lem:beta} and $m +1 \geq 2\K^{(x,z)}/\ge$ for the last inequality. Rearrange \eqref{eq:c-secondhalf} and sum with \eqref{eq:c-firsthalf} to get \eqref{eq:suffices}. \qed
\\

\noindent {\bf Proof of \propref{prop:bes:etal:2008:1}.} This follows immediately from Propositions \ref{prop:qm} and \ref{prop:lattice}. It can also be viewed as an application of Proposition 1 of \citet{bes:etal:2008} to a function $g(u_1,u_2,u_3) \propto u_1u_2/u_3$. \qed
\\

\noindent {\bf Proof of \propref{prop:neutralcomplexity}.} 
(i) Using \eqref{eq:ratio} and \eqref{eq:finequality},
\begin{equation}
\label{eq:finequality2}
\frac{b^{(t,\theta)}_{K+1}(m)}{b^{(t,\theta)}_{K}(m)} = f^\theta_m(K)e^{-\frac{(2K+\theta)t}{2}} < (\gq+2K+1)e^{-\frac{(2K+\theta)t}{2}}.
\end{equation}
The right-hand side of \eqref{eq:finequality2} is less than 1 provided
\[
K > \frac{\log(\theta + 2m + 1)}{t} - \frac{\theta}{2},
\]
which implies that $\Ca_m^{(t,\gq)} < \frac{\log(\theta + 2m + 1)}{t} - \frac{\theta}{2} = O(t^{-1})$ as $t\to 0$.\\
(ii) Continuing,
\[
\Ca_m^{(t,\gq)} < \frac{\log(\theta + 2m + 1)}{t} - \frac{\theta}{2} < \frac{\log(\theta + 2\Cb^{(t,\gq)}_0 + 1)}{t} - \frac{\theta}{2},
\]
with the right-hand side independent of $m$ and asymptotically $O(t^{-1}\log(t^{-1}))$ as $t\to 0$ by (iii) below.\\
(iii) By inspection of the definition \eqref{eq:C} of $\Cb_0^{(t,\gq)}$, in order to ensure $K > \Cb_0^{(t,\gq)}$ it suffices that
\[
K > \frac{1 \vee \log(\theta + 2K + 1)}{t},
\]
which holds for sufficiently small $t$ if $K \sim t^{-(1+\gk)}$, for any fixed $\gk > 0$. Hence, $\Cb^{(t,\gq)}_0 = o(t^{-(1+\gk)})$ as $t\to 0$.\\
(iv) Parts (i--iii) cover those terms that must be precomputed in Algorithm \ref{alg:q}: For the random number of remaining terms we may assume $K \geq  m+\Ca_m^{(t,\gq)}$, so that $b_{K+1}(m) < b_K(m)$. Write $Q_M^{\gq}(t) := \bbP(A^\theta_\infty(t) \leq M)$. Our aim is to show that these random remaining terms do not add to the complexity of the calculation beyond (iii); we achieve this by determining the complexity of
\[
\Cc_\gd^{(t,\gq)}(M) := \inf\left\{ K \geq \max_{m\in\{0,\dots,M\}}(m+\Ca_m^{(t,\gq)}) : \sum_{m=0}^M \sum_{k=K}^\infty (-1)^{k-m} b_k(m) < \gd \right\}
\]
as $t\to 0$, for a fixed $\gd > 0$. This is the appropriate quantity to look at, since if $K \geq \Cc_\gd^{(t,\gq)}(M)$ then $|Q_M^{\gq}(t) - S^{\pm}_\bfk(M)| < \gd$, when $2k_i \geq K$; $i=0,\dots,M$. 
 Furthermore, averaging over the uniform random variable $U$ drawn for inversion sampling, the total number of coefficients $N^{(t,\gq)}\mid U = u$ needed to determine that $Q_{m-1}^{\gq}(t) < u < Q_m^{\gq}(t)$ is given by the number of terms needed to bound both our estimates of $Q_{m-1}^{\gq}(t)$ and $Q_m^{\gq}(t)$ away from $u$:
\begin{align}
\bbE(N^{(t,\gq)}) &= \bbE[\bbE(N^{(t,\gq)}\mid U)] \notag\\
&\leq \sum_{m=0}^\infty \int_{Q_{m-1}^{\gq}(t)}^{Q_{m}^{\gq}(t)}  [\Cc_{Q_{m}^{\gq}(t)-u}^{(t,\gq)}(m) +  \Cc_{u-Q_{m-1}^{\gq}(t)}^{(t,\gq)}(m-1)] du. \label{eq:EN}
\end{align}
We will show that if $K \sim t^{-(1+\gk)}$ as $t\to 0$, for a fixed $\gk > 0$, then we can attain the stated growth condition on $\bbE(N^{(t,\gq)})$. 
Using that the right-hand side of \eqref{eq:finequality2} is decreasing in $K$ for $K \geq m+\Ca_m^{(t,\gq)}$, for each $\gz < 1$ we can find a constant $c_1^{(\gq)}$ such that for $K > c_1^{(\gq)}t^{-1}$, $f_m^\gq(K)e^{-(2K+\gq)t/2} < \gz$. Hence,
\begin{multline}
\sum_{m=0}^M \sum_{k=K}^\infty (-1)^{k-m} b_k(m) < \sum_{m=0}^M \sum_{k=K}^\infty b_k(m) \\
<  \sum_{m=0}^M \sum_{k=K}^\infty \gz^{k-K} b_K(m) = \sum_{m=0}^M \frac{b_K(m)}{1-\gz}. \label{eq:complexity}
\end{multline}
Routine calculations show that, for $m \geq 1$,
\begin{multline*}
a^\gq_{k,m+1} = a^\gq_{km} \frac{(k-m)(\gq+m+k-1)}{(m+1)(\gq+m)} \\\leq a^\gq_{km} \frac{(k-1)(\gq+k)}{2(\gq+1)} \leq a^\gq_{k1}\left[\frac{(k-1)(\gq+k)}{2(\gq+1)}\right]^{k-1},
\end{multline*}
and thus, applying Stirling's formula to $a_{k1}^\gq \sim k^{c_2^{(\gq)}}$, 
\begin{equation}
\label{eq:complexity2}
b_K(m) = a^\gq_{Km}e^{-K(K+\gq-1)t/2} \leq e^{c_3^{(\gq)}K\log K - K(K+\gq-1)t/2} \leq c_4^{(\gq)}e^{-K^2t/2},
\end{equation}
for some constants $c_2^{(\gq)}, c_3^{(\gq)}$, $c_4^{(\gq)}$ (which again exist by our assumption about the asymptotic growth of $K$). In the special case $m=0$, Stirling's formula also yields $a^\gq_{k0} \sim k^{c_2^{(\gq)}}$ and so the inequalities in \eqref{eq:complexity2} continue to hold. 
Combining \eqref{eq:complexity} with \eqref{eq:complexity2} we find
\[
\sum_{m=0}^M \sum_{k=K}^\infty (-1)^{k-m} b_k(m) < c_5^{(\gq)}(M+1)e^{-K^2t/2},
\]
for some $c_5^{(\gq)}$, which is less than $\gd$ if
\begin{equation}
\label{eq:complexity3}
K > \sqrt{\frac{2}{t}}\log\left(\frac{c_5^{(\gq)}(M+1)}{\gd}\right).
\end{equation}
(This is not a tight bound but suffices in the calculations below.) In summary, if $K$ satisfies both $K > c_1^{(\gq)}t^{-1}$ and \eqref{eq:complexity3} then $K > \Cc_\gd^{(t,\gq)}(M)$. Integrating over $\gd$:
\begin{multline}
\int_{Q_{m-1}^{\gq}(t)}^{Q_{m}^{\gq}(t)}  \Cc_{Q_{m}^{\gq}(t) -u}^{(t,\gq)}(m) du = \int_{0}^{q_{m}^{\gq}(t)}  \Cc_{\gd}^{(t,\gq)}(m) d\gd \\
< \int_{0}^{q_{m}^{\gq}(t)} \left[\frac{c_1^{(\gq)}}{t^{1+\gk}} + \sqrt{\frac{2}{t}}\log\left(\frac{c_5^{(\gq)}(m+1)}{\gd}\right)\right]  d\gd\\
= q_{m}^{\gq}(t)\left[\frac{c_1^{(\gq)}}{t^{1+\gk}} + \sqrt{\frac{2}{t}}\left[\log(c_5^{\gq}(m+1)) + 1 - \log q_{m}^{\gq}(t)\right]\right). \label{eq:complexitybound}
\end{multline}
It remains to show that the resulting expression \eqref{eq:complexitybound} can be summed over $m$, which is possible by \thmref{thm:gri:1984}; in particular, 
\begin{align*}
q_{m}^{\gq}(t) &= \lefteqn{\frac{1}{\sqrt{2\pi (\gs^{(t,\gq)})^2}}\exp\left(\frac{(m - \mu^{(t,\gq)})^2}{2(\gs^{(t,\gq)})^2}\right) + o(1),}\\
\mu^{(t,\gq)} &= \frac{2}{t} + O(1), & (\gs^{(t,\gq)})^2 &= \frac{2}{3t} + O(1).
\end{align*}
Hence, by Jensen's inequality,
\begin{align*}
\sum_{m=0}^\infty q_m^\gq(t) \log m &= \bbE[\log A_\infty^\gq (t)] \leq \log \bbE[A_\infty^\gq(t)] = O(\log t^{-1}),\\
\sum_{m=0}^\infty q_m^\gq(t) [-\log q_m^\gq (t)] &= \bbE\left[\log (\sqrt{2\pi}\gs^{(t,\gq)}) + \frac{1}{2}\left(\frac{A_\infty^\gq (t)-\mu^{(t,\gq)}}{\gs^{(t,\gq)}}\right)^2\right]\\
&= O(\log t^{-1}) + \frac{1}{2}\bbE[X^2] + O(1) = O(\log t^{-1}),
\end{align*}
where $X \sim N(0,1)$. Combining these results with \eqref{eq:complexitybound}, we obtain
\begin{align*}
\sum_{m=0}^\infty \int_{Q_{m-1}^{\gq}(t)}^{Q_{m}^{\gq}(t)}  \Cc_{Q_{m}^{\gq}(t) -u}^{(t,\gq)}(m) du &= O(t^{-(1+\gk)}) + O(t^{-1/2}\log t^{-1})\\
&= O(t^{-(1+\gk)}),
\end{align*}
 showing that the first term on the right of \eqref{eq:EN} is $O(t^{-(1+\gk)})$. The second term is also $O(t^{-(1+\gk)})$ by a similar argument. Since $\gk$ was arbitrary, $\bbE[N^{(t,\gq)}] = o(t^{-(1+\gk)})$. \qed
\\

\noindent {\bf Proof of \propref{prop:coopbob}.}
For a diffusion with drift $\gamma$, since $\eta$ is continuous on $(0,1)$, then
$$\int_0^T \frac{\gamma^2(X_t)-\alpha^2(X_t)}{X_t(1-X_t)}\ dt=\int_0^T\ \left[\eta^2(X_t)X_t(1-X_t)+2\eta(X_t)\alpha(X_t)\right]\ dt<\infty,$$
then Novikov's condition is satisfied and a Girsanov transform exists with respect to the neutral WF diffusion with drift $\alpha$, i.e.\ (WF1) holds. In particular,  \eqref{eq:GirsanovWF} reads
\begin{equation}
\exp\left\{\int_0^T\eta(X_t)\ dX_t-\frac{1}{2}\int_0^T\ \left[\eta^2(X_t)X_t(1-X_t)+2\eta(X_t)\alpha(X_t)\right]\ dt\right\}.
\label{eq:cggirsanov}
\end{equation}
The function $\eta$ is also continuously differentiable in $(0,1)$, so $\gamma^{\prime}(x)-\alpha^{\prime}(x)=\eta^{\prime}(x)x(1-x)+\eta(x)(1-2x)$ is continuous and
\begin{equation}
\widetilde\phi(x)=\frac{1}{2}\left[x(1-x)(\eta^2(x)+\eta^{\prime}(x))+2\eta(x)\alpha(x) \right]
\label{eq:cgepigraph}
\end{equation}
 which, being itself continuous in $[0,1]$, is then bounded in $(0,1),$ and (WF3) follows.\\
(WF2) and (WF4) are obvious. Thus \eqref{eq:cggirsanov} has a version of the form \eqref{eq:GirsanovWF2} with $\widetilde\phi$ as in \eqref{eq:cgepigraph} and $\widetilde A(x)=\int_0^x\eta(z)\ dz$. \qed

\section*{Acknowledgements}
We thank Alison Etheridge, Bob Griffiths, Jere Koskela, Thomas Kurtz, Omiros Papaspiliopoulos, and Murray Pollock for useful discussions, and the anonymous referees for numerous helpful comments.

% BibTeX support
%\newpage
\bibliographystyle{imsart-nameyear}
\bibliography{master}
%{\sc
%\begin{multicols}{2}
%\noindent Paul A.\ Jenkins\\
%Department of Statistics\\
%University of Warwick\\
%Coventry CV4 7AL, U.K.\\
%{\sc E-mail}: p.jenkins@warwick.ac.uk
%
%\noindent Dario Span\`o\\
%Department of Statistics\\
%University of Warwick\\
%Coventry CV4 7AL, U.K.\\
%{\sc E-mail}: d.spano@warwick.ac.uk
%\end{multicols}
%}
\end{document}